\documentclass[preprint,12pt]{elsarticle}

\usepackage{subcaption}
\usepackage{amssymb}
\usepackage{booktabs}
\usepackage{amsmath}
\usepackage{graphicx}
\usepackage{multirow}
\usepackage[backref]{hyperref}

\usepackage[linesnumbered,boxed,ruled,commentsnumbered]{algorithm2e}
\usepackage{color,xcolor} 

\usepackage{tabularx}
\usepackage{hyperref}
\usepackage[figuresright]{rotating}

\begin{document}

\begin{frontmatter}



\title{Trend-encoded Probabilistic Multi-Order Model: A Non-Machine Learning Approach for Enhanced Stock Market Forecasts}

  \author[mymainaddress]{Peiwan Wang\fnref{fn1}}\fntext[fn1]{Corresponding author.}
		\ead{peiwanwang@ustc.edu.cn}
        \author[mysecondaryaddress]{Chenhao Cui}
        \ead{niklaus@mail.ustc.edu.cn} 
        \author[address3]{Yong Li}\ead{yonglee@ustc.edu.cn}
     
        \address[mymainaddress]{School of Mathematics and Big Data, Chaohu University,\\ 1 Bantang Road, Chaohu Economic Development Zone, Hefei City, Anhui Province, P.R.China, 238024}
        \address[mysecondaryaddress]{Smart Earth Sensing (Hefei) Technology Co., Ltd., \\ 9 Wangjiang West Road, Hefei City, Anhui Province, P.R.China, 230031}
        \address[address3]{New Finance Research Center, International Institute of Finance, University of Science and Technology of China,\\ 1789 Guangxi Road, Hefei City, Anhui Province, P.R.China, 230092}

\begin{abstract}
In recent years, the dominance of machine learning in stock market forecasting has been evident. While these models have shown decreasing prediction errors, their robustness across different datasets has been a concern. A successful stock market prediction model not only minimizes prediction errors but also showcases robustness across various data sets, indicating superior forecasting performance. This study introduces a novel multiple lag order probabilistic model based on trend encoding (TeMoP) that enhances stock market predictions through a probabilistic approach. Results across different stock indexes from nine countries demonstrate that the TeMoP outperforms the state-of-art machine learning models in terms of predicting accuracy and stabilization. 
\end{abstract}


\end{frontmatter}

\section{Introduction}\label{sec:intro}

  Time series prediction is an evergreen topic that spans across various application domains. For instance, it finds applications in fields such as economics \citep{masini2023machine}, finance \citep{CUI2023120902}, traffic flow \citep{lv2014traffic}, and energy \citep{wang2017deep}, among others. Time series trend forecasting serves as the foundation of time series prediction. It involves using historical data to forecast whether future data will exhibit an upward trend. Accurately capturing the trend of a time series is essential for achieving precise time series forecasts. Therefore, designing a model with superior forecasting performance for time series trend forecasting holds significant value in practical applications.
  
  In practical applications, a good stock market prediction model not only has a small prediction error, but this small prediction error is also well robust across different data sets. However, existing models often struggle to meet these two characteristics at the same time. Existing time series trend forecasting models can be broadly categorized into four types based on their methodological principles: statistical models \citep{RAY2023110939}, machine learning models in a narrow sense \citep{costa2023recent,yin2023research}, deep learning models \citep{9120214}, and fuzzy time series models \citep{BIteNCOURT2023127072}. Compared to the other models, deep learning models and fuzzy time series models demonstrate less prediction errors on the test set. However, poor robustness and black-box nature of deep learning models and the information leakage issue of fuzzy time series models have been criticized.

 Therefore, for time series trend forecasting, this paper proposes a multiple lag order probabilistic model based on trend encoding (TeMoP) inspired by fuzzy time series models and random forests. In fuzzy time series modeling, time series forecasting is achieved by blurring the time series into different degrees and analyzing the patterns between the different degrees. Therefore, in addition to using the original features, this paper also obtains the trend features of the time series through trend encoding and analyzes the patterns between different trend features, which in turn reduces the prediction errors. In random forests, several different models are trained based on different bootstrap sets, and then the results of these models are combined thereby improving the robustness of predictive performance. Influenced by random forests, to solve the problem of poor robustness of models trained by samples with a certain lag order, this paper trains different models based on samples with different lag orders, and then combines the results of these models to improve the robustness of prediction performance.
  
  Thus, in principle, TeMoP offers the following advantages: (1) Excellent robustness. Unlike classical models trained by samples with a certain lag order, TeMoP integrates the results of models trained by samples with different lag order, which enhances robustness. (2) Small prediction errors. The introduction of trend features reduces the prediction errors of TeMoP.


  Due to its inherent characteristics such as high volatility and non-normality, stock data is renowned for being difficult to predict. However, effective prediction of stock data is crucial for providing sound investment decision recommendations. Therefore, this study selects the stock trend prediction as the application scenario. By analyzing stock data from different types of financial markets, it indicates that TeMoP outperforms the comparative models in terms of prediction errors. Furthermore, unlike the comparative models whose prediction errors fluctuates wildly across different data sets, TeMoP demonstrates good robustness.

  The main innovations and contributions of this paper are as follows:
  
  \begin{itemize}
      \item This study proposes a multiple lag order probabilistic model based on trend encoding (TeMoP) for forecasting time series trends. In principle, this model shows good robustness and small prediction errors.
      \item Building upon the background of stock trend prediction, this study analyzes the superiority of TeMoP in terms of prediction errors and robustness across multiple data sets from various types of financial markets.
  \end{itemize} 
  
  The remaining sections of this paper are structured as follows: The second section \ref{sec:Lr} reviews and analyzes related research. The third section \ref{sec:method} provides a detailed introduction to the principles of TeMoP. The fourth section \ref{sec:data_results} conducts a comprehensive comparative analysis of the experimental results. The fifth section \ref{sec:sum} summarizes the entire paper and provides outlooks for future research. The last section consists of the references.

\section{Literature review}\label{sec:Lr}

  For a long time, research on time series trend forecasting has been abundant. Based on different methodological principles, existing research results can be classified into four categories: statistical models (SM), machine learning models in a narrow sense (ML), deep learning models (DL), and fuzzy time series models (FTSM).

  The complexity of the real world often leads to statistical models failing to meet their assumptions, resulting in poor prediction performance on real-world data sets \citep{LI2023110489,9667230,WANG20231163}. Unlike statistical models, machine learning models are data-driven. For example, Random Forests, a classic model in machine learning, improve both data utilization efficiency and robustness by sampling from the original data set using bootstrapping and integrating the output of multiple models \citep{JAIN2023119225,sun2024improved,HE2024120478}. Deep learning methods are also data-driven. Despite small prediction errors on real-world data sets, they are criticized for poor robustness \citep{9210118,9942340,9868261}. Unlike the previous models, fuzzy time series models are designed based on fuzzy mathematics principles \cite{silva2020distributed}. Fuzzy time series modeling mainly involves: (1) defining the universe of discourse and linguistic variables, (2) fuzzification, (3) extracting fuzzy temporal patterns, (4) building a fuzzy rule base, (5) making predictions based on the fuzzy rule base, and (6) defuzzification.

  Among numerous researches, fuzzy time series models and deep learning-based decomposition-ensemble models exhibit less prediction errors on test sets. Compared to deep learning, fuzzy time series models constitute a niche research direction. Existing researches mainly fall into two categories. One is innovative research on the theoretical methods of fuzzy time series models \citep{10432971,wang2024bayesian}. For instance, \cite{de2019probabilistic} proposes probabilistic weighted fuzzy time series model (PWFTS), which not only integrates point, interval, and probabilistic forecasting in the same model, but also is a new non-parametric, data driven, and highly accurate forecasting method. The other is innovative research on the practical applications of fuzzy time series models \citep{sadaei2019short,severiano2021evolving}. For example, \cite{bitencourt2023combining} combines Autoencoder with weighted multivariate fuzzy time series model (WMVFTS) \citep{de2019distributed} for high-dimensional time series forecasting in the Internet of Energy (IoE), and the proposed model achieves excellent prediction performance. Through extensive comparative analysis, these researches are all based on an open-source Python library---pyfts \citep{silvapyfts}. However, a study of the pyfts source code reveals a flaw in information leakage during the prediction process. That is, pyfts utilizes future information data during prediction, resulting in excellent performance of the prediction results based on pyfts. 
  
  Similarly, the deep learning-based decomposition-ensemble method also suffers from information leakage \citep{he2022modeling,HE2021107488}. The main steps of this method are as follows: (1) decompose the original time series into multiple sub-sequence using sequence decomposition algorithms, (2) model and predict each sub-sequence separately, and (3) sum the predictions of each sub-sequence to obtain the final prediction result. Common sequence decomposition algorithms include Seasonal and Trend decomposition using Loess (STL) \citep{HE2021107488,zeng2024novel}, Empirical Mode Decomposition (EMD) \citep{JUN2017167,XIANG2018874}, Complete EEMD with Adaptive Noise (CEEMDAN) \citep{LI2024131448,SONG2024111729}, and Variational Mode Decomposition (VMD) \citep{NASIRI2023110867,ZHANG2023119617}. This method often exhibits small prediction errors on the test set. For instance, \cite{liang2022forecasting} first decomposes gold prices into multiple sub-sequences using CEEMDAN, then combines Long Short term Memory (LSTM), Convolutional Neural Networks (CNN), and Convolutional Block Attention Module (CBAM) to predict each sub-sequence, and finally sums the prediction results of all sub-sequences to obtain the final prediction results. Experimental results shows that, on the test set, the proposed model achieves a determination coefficient (R$^2$) exceeding $0.999$ and Mean Absolute Percentage Error (MAPE) below $0.3\%$. In the method, the test set, which should be unknown, is treated as known, thus inevitably causing the leakage of future information. 
  
  In summary, although fuzzy time series models and deep learning-based decomposition-ensemble methods exhibit small prediction errors on the test set, their flaws in information leakage render these models lack practical significance.

  Therefore, based on the shortcomings of existing models, this study designs a multiple lag order probabilistic model based on trend encoding TeMoP for time series trend prediction, which performs excellently in terms of robustness, prediction errors, and other aspects from the perspective of practical application. The next section \ref{sec:method} will focus on introducing the principles of TeMoP.

\section{Methodology}\label{sec:method}

\subsection{preliminary}

The symbols and their meanings in this paper are as follows:
\begin{itemize}
    \item $Z$: Time series data for training. $Z=\{z_1,z_2,...,z_{n-1}\}$.
    \item $m$: The threshold for determining whether a sample set is a large sample. Typically, in a statistical context, $m=50$.
    \item $q$: Maximum lag order of samples.
    \item $x^i_j$: A sample with lag order $i$. $x^i_j=(z_{j_{n-i}},z_{j_{n-i+1}}...z_{j_{n-1}})$.
    \item $y^i_j$: A sample label with lag order $i$. $y^i_j = (z_{j_n})$.
    \item $\Omega^i$: A set of samples with lag order $i$. $\Omega^i = \{x^i_j| j=1,2,...,n-i-2\}$.
    \item $t(x^i_j)$: Trend encoding function for samples $x^i_j$.
    \item $\Omega^i_l$: A subset of $\Omega^i$ and all samples in this subset exhibit identical trend encoding results. $\Omega^i_l \subseteq  \Omega^i, \forall x^i_j,x^i_h \in \Omega^i_l, t(x^i_j) = t(x^i_h)$.
    \item $t(\Omega^i_l)$: Trend feature vectors of $\Omega^i_l$. $t(\Omega^i_l)=\{t(x^i_j)| \forall x^i_j \in \Omega^i_l\}$.
    \item $+\Omega^i_l$: A subset of $\Omega^i_l$ and trend encoding results of sample label corresponding to samples in this subset are equal to $1$. $+\Omega^i_l \subseteq \Omega^i_l, \forall x^i_j \in +\Omega^i_l, t(y^i_j)=1 $.
    \item $-\Omega^i_l$: A subset of $\Omega^i_l$ and trend encoding results of sample label corresponding to samples in this subset are equal to $-1$. $-\Omega^i_l \subseteq \Omega^i_l, \forall x^i_j \in -\Omega^i_l, t(y^i_j)=-1 $.
    \item $\boldsymbol{\mu}(\Omega)$: Sample mean of set $\Omega$.
    \item $\boldsymbol{\Sigma}(\Omega)$: Sample covariance of set $\Omega$.
    \item $|\Omega|$: Number of all elements in set $\Omega$.
    \item $N(\Omega)$: Normalization function in set $\Omega$. 
    \item $\hat{Z}$: Time series data for test. $\hat{Z}=\{\hat{z}_{n-q},\hat{z}_{n-q+1},...,\hat{z}_{n-1}\}$.
    \item $U_{\Omega^i_l}(x^i_h)$: Similarity of $t(x^i_h)$ and $t(\Omega^i_l)$.
    \item $S(t(y^i_0)=1 | x^i_0 \in \Omega^i_l, t(x^i_0))$: Score of $t(y^i_0)=1$ computed based on $t(x^i_0)$ when $x^i_0 \in \Omega^i_l$.
    \item $S(t(y^i_0)=1 | x^i_0 \in \Omega^i_l, x^i_0)$: Score of $t(y^i_0)=1$ computed based on $x^i_0$ when $x^i_0 \in \Omega^i_l$.
    \item $S(t(y^i_0)=1 |t(x^i_0))$: Score of $t(y^i_0)=1$ computed based on $t(x^i_0)$.
    \item $S(t(y^i_0)=1 |x^i_0)$: Score of $t(y^i_0)=1$ computed based on $x^i_0$.
    \item $S(t(y^i_0)=1)$: Score of $t(y^i_0)=1$.
\end{itemize}

There are some functions in this paper and their meanings and mathematical formulas are as follows.

\begin{itemize}
    \item $t(z_i)$ \ref{functions: Trend encoding function for scalar} denotes trend encoding function for scalar. For vectors, its trend encoding function is $t(z_{i_1},z_{i_2},...,z_{i_p}) = (t(z_{i_2}),t(z_{i_3}),...,t(z_{i_p}))$.

    \begin{equation}\label{functions: Trend encoding function for scalar}
        t(z_i)=\left\{
    	\begin{aligned}
    	-1, \quad z_i<z_{i-1},\\
    	1, \quad z_i \geq z_{i-1},\\
    	\end{aligned}
    	\right.
    \end{equation}

    \item $I[x_i = y_i]$ denotes indicator function and the formula is as follows.

    \begin{equation}
        I[x_i = y_i]=\left\{
    	\begin{aligned}
    	1, \quad x_i = y_i,\\
    	0, \quad others.\\
    	\end{aligned}
    	\right.
    \end{equation}

    \item $x \vec{\wedge} y$ \ref{functions: Overlap of vectors} denotes overlap of vectors.

    \begin{equation}\label{functions: Overlap of vectors}
        x \vec{\wedge} y = \frac{1}{n} \sum^n_{i=1} I[x_i = y_i]
    \end{equation}

    \item $M(x^i_j, +\Omega^i_l)$ \ref{functions: Mahalanobis Distance} denotes mahalanobis distance between $x^i_j$ and $+\Omega^i_l$.

    \begin{equation}\label{functions: Mahalanobis Distance}
        M(x^i_j, +\Omega^i_l)=\sqrt{(x^i_j - \mu(+\Omega^i_l))^T \Sigma(+\Omega^i_l)^{-1} (x^i_j - \mu(+\Omega^i_l))}
    \end{equation}

    \item $ Dtp(u) $ is a mapping the mahalanobis distance to a value between $[0,1]$.

    \begin{equation}\label{functions: mapping}
        Dtp(u) = \frac{2}{1+e^u}
    \end{equation}
\end{itemize}

\begin{figure}[!ht]
\centering
\includegraphics[scale=0.5]{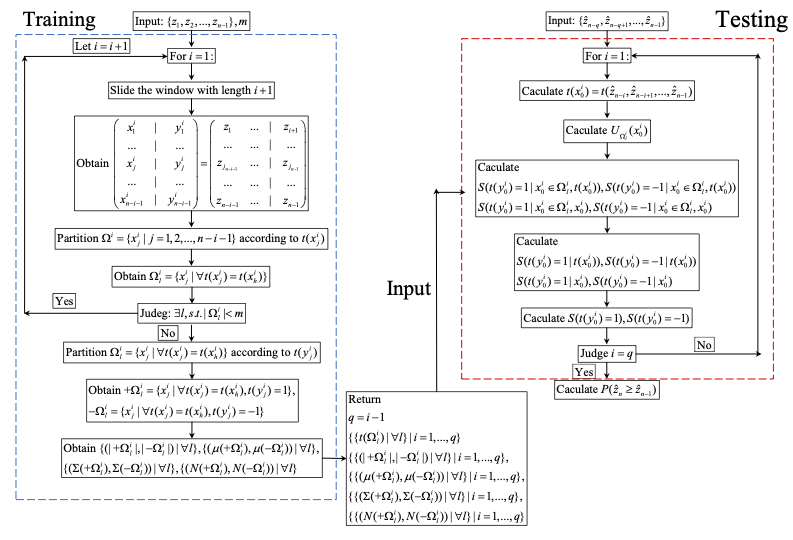}
\caption{Algorithm of TeMoP.}\label{fig:temop}
\end{figure}

Figure \ref{fig:temop} is flowchart of the algorithm for TeMoP. According to the Figure \ref{fig:temop}, TeMoP is mainly divided into a training section and a testing section. $Z=\{z_1,z_2,...,z_{n-1}\}$ is time series data for training. And $\hat{Z}=\{\hat{z}_{n-q},\hat{z}_{n-q+1},...,\hat{z}_{n-1}\}$ is test data for calculating $P( \hat{z}_n \geq \hat{z}_{n-1})$ with $\hat{z}_n$ unknown.

\subsection{Training}


For time series prediction, traditional models firstly have to determine lag order of samples. Unlike the traditional models, TeMoP starts training from the lag order equal to $1$, and determines the value of maximum lag order adaptively by setting the judgment condition and nature of data. The training process is as follows.

\begin{itemize}
      \item[\textbf{1:}] Let $i=1$.
      \item[\textbf{2:}] Slide the windows with length $i+1$ to obtain samples and labels like Equation \ref{equation:samples_labels}.
      
        \begin{equation}\label{equation:samples_labels}
            \begin{pmatrix}
            x^i_1 & \vdots & y^i_1 \\
            x^i_2 & \vdots & y^i_2 \\
            ... &  & ... \\
            x^i_{n-i-1} & \vdots & y^i_{n-i-1} \\
        \end{pmatrix} = \begin{pmatrix}
            z_1 & ... & z_i & \vdots & z_{i+1} \\
            z_2 & ... & z_{i+1} & \vdots & z_{i+2} \\
            ... & ... & ... &   & ... \\
            z_{n-i-1} & ... & z_{n-2} & \vdots & z_{n-1} \\
        \end{pmatrix}
        \end{equation}

      \item[\textbf{3:}] Let $\Omega^i = \{x^i_j,j=1,2,...,n-i-1\}$. Then split $\Omega^i$ to obtain $\Omega^i_l = \{x^i_j| \forall t(x^i_j) = t(x^i_h)\}$ according to $t(x^i_j)$ and return $\{t(\Omega^i_l)| \forall l\}$. 
      \item[\textbf{4:}] Judge whether there exists $l$ such that $|\Omega^i_l|<m$. If yes, then stop and let $i=i+1$ and repeat steps $2$ to $4$. If no, then continue next steps.
      \item[\textbf{5:}] Split $\Omega^i_l = \{x^i_j| \forall t(x^i_j) = t(x^i_h)\}$ according to $t(y^i_j)$. Let $+\Omega^i_l = \{x^i_j| \forall t(x^i_j) = t(x^i_h),t(y^i_j)=1\}$ and $-\Omega^i_l = \{x^i_j| \forall t(x^i_j) = t(x^i_h),t(y^i_j)=-1\}$.
      \item[\textbf{6:}] Compute and return 
      \begin{align*}
          &\{(|+\Omega^i_l|,|-\Omega^i_l|) | \forall l\},\\
          &\{(\mu(+\Omega^i_l),\mu(-\Omega^i_l)) | \forall l\},\\
          &\{(\Sigma(+\Omega^i_l),\Sigma(-\Omega^i_l)) | \forall l\},\\
          &\{(N(+\Omega^i_l),N(-\Omega^i_l)) | \forall l\}.
      \end{align*}
      
      \item[\textbf{7:}] Return
      \begin{align*}
          &q=i-1,\\
          &\{\{t(\Omega^i_l)| \forall l\} | i=1,...,q\},\\
          &\{\{(|+\Omega^i_l|,|-\Omega^i_l|) | \forall l\} | i=1,...,q\},\\
          &\{\{(\mu(+\Omega^i_l),\mu(-\Omega^i_l)) | \forall l\} | i=1,...,q\},\\
          &\{\{(\Sigma(+\Omega^i_l),\Sigma(-\Omega^i_l)) | \forall l\} | i=1,...,q\},\\
          &\{\{(N(+\Omega^i_l),N(-\Omega^i_l)) | \forall l\} | i=1,...,q\}.
      \end{align*}
  \end{itemize}

\subsection{Testing}

In this section, based on results of training, we use $\{\hat{z}_{n-q},\hat{z}_{n-q+1},...,\hat{z}_{n-1}\}$ to compute $P( \hat{z}_n \geq \hat{z}_{n-1})$ with $\hat{z}_n$ unknown. The calculation process is as follows.

\begin{itemize}
      \item[\textbf{1:}] Let $i=1$.
      \item[\textbf{2:}] Let $x^i_0 = \{\hat{z}_{n-i},\hat{z}_{n-i+1},...,\hat{z}_{n-1}\}, y^i_0 = \hat{z}_n$ and compute $t(x^i_0)$.
      \item[\textbf{3:}] Compute $U_{\Omega^i_l}(x^i_0)$.
      
      \begin{equation}
        U_{\Omega^i_l}(x^i_0)=t(x^i_0) \vec{\wedge} t(\Omega^i_l).
      \end{equation}
      
      \item[\textbf{4:}] Compute $S(t(y^i_0)=1 | x^i_0 \in \Omega^i_l, t(x^i_0))$ and $S(t(y^i_0)=1 | x^i_0 \in \Omega^i_l, x^i_0)$ based on Equation \ref{equation:trend_score1} and Equation \ref{equation:ori_score1} respectively. $S(t(y^i_0)=-1 | x^i_0 \in \Omega^i_l, t(x^i_0))$ and $S(t(y^i_0)=-1 | x^i_0 \in \Omega^i_l, x^i_0)$ ditto.
      
      \begin{equation}\label{equation:trend_score1}
          S(t(y^i_0)=1 | x^i_0 \in \Omega^i_l, t(x^i_0)) = \frac{|+\Omega^i_l|+1}{|+\Omega^i_l|+|-\Omega^i_l|+2}
      \end{equation}

      \begin{equation}\label{equation:ori_score1}
          S(t(y^i_0)=1 | x^i_0 \in \Omega^i_l, x^i_0) = Dtp(M(x^i_0, +\Omega^i_l))
      \end{equation}
      
      \item[\textbf{5:}] Compute $S(t(y^i_0)=1 | t(x^i_0))$ and $S(t(y^i_0)=1 |x^i_0)$ based on Equation \ref{equation:trend_score2} and Equation \ref{equation:ori_score2} respectively. $S(t(y^i_0)=-1 | t(x^i_0))$ and $S(t(y^i_0)=-1 |x^i_0)$ ditto.
      
        \begin{equation}\label{equation:trend_score2}
            S(t(y^i_0)=1 | t(x^i_0)) = \sum_l U_{\Omega^i_l}(x^i_0) \cdot S(t(y^i_0)=1 | x^i_0 \in \Omega^i_l, t(x^i_0)),
        \end{equation}
        
        \begin{equation}\label{equation:ori_score2}
            S(t(y^i_0)=1 | x^i_0) = \sum_l U_{\Omega^i_l}(x^i_0) \cdot S(t(y^i_0)=1 | x^i_0 \in \Omega^i_l, x^i_0).
        \end{equation}
      
      \item[\textbf{6:}] Compute $S(t(y^i_0)=1)$ based on Equation \ref{equation:all_score}. $S(t(y^i_0)=-1)$ ditto.
        \begin{equation}\label{equation:all_score}
            S(t(y^i_0)=1) = S(t(y^i_0)=1 | x^i_0) + S(t(y^i_0)=1 | t(x^i_0)).
        \end{equation}

      \item[\textbf{7:}] Judge whether $i=q$. If yes, compute $P( \hat{z}_n \geq \hat{z}_{n-1})$ based on Equation \ref{eqaution:proba}. If no, let $i=i+1$ and repeat steps $2$ to $7$.
        \begin{equation}\label{eqaution:proba}
            P( \hat{z}_n \geq \hat{z}_{n-1}) = P(t(\hat{z}_n)=1)=\frac{e^{\sum_{i=1}^q S(t(y^i_0)=1)}}{e^{\sum_{i=1}^q S(t(y^i_0)=1)}+e^{\sum_{i=1}^q S(t(y^i_0)=-1)}}.
        \end{equation}
      
\end{itemize}

\textbf{Caution:} Samples need to be normalized before calculating the Mahalanobis distance. After normalization, the $\Sigma(+\Omega^i_l)$ takes values close to the unit matrix $E$, which leads to large values of $\Sigma(+\Omega^i_l)^{-1}$ and further leads to large computed Mahalanobis distances. Therefore, in order to solve the above issue, $\Sigma(+\Omega^i_l) + 0.1 \cdot E$ is used to instead of $\Sigma(+\Omega^i_l)$. 

\subsection{Why}

The above subsections answers how TeMoP can be utilized for training and testing, and this section provides a detailed explanation of why TeMoP is so designed.

\subsubsection{The non-stationary nature of the data}

In the real world, time series data often exhibit non-stationary properties. Therefore, models trained based on samples with a certain lag order tend to exhibit poor robustness.

Therefore, in order to solve the above problem, this paper firstly computes the maximum lag order by the nature of the data. Then, this paper synthesize results of various models trained based on samples with different lag order not exceeding the maximum lag order, as shown in Equation \ref{eqaution:proba}.

\subsubsection{Hypothesis: The trend features of the sample significantly impact the performance of time series trend forecasting.}

In Equation \ref{eqaution:proba}, $S(t(y^i_0)=1)$ denotes score of $t(y^i_0)=1$ based on samples with lag order $i$. In this paper, it is argued that the trend features of samples are also important factors in the time series trend forecasting problem. Therefore, this paper argues that the score of $t(y^i_0)=1$ when the lag order of samples is $i$ is equal to the score calculated based on original features $x^i_0$ plus the score calculated based on the trend features $t(x^i_0)$, as shown in Equation \ref{equation:all_score}.

\subsubsection{Partition of the sample space}

When the lag order of samples is $i$, $\Omega^i = \{x^i_j,j=1,2,...,n-i-1\}$ is naturally divided into different subsets $\Omega^i_l = \{x^i_j| \forall t(x^i_j) = t(x^i_h)\}$ based on the trend features of samples. And $\cup_l \Omega^i_l = \Omega^i$, $\forall k \neq l, \Omega^i_l \cap \Omega^i_k = \varnothing$.

Then, $\Omega^i_l$ are naturally partitioned into two different subsets $+\Omega^i_l = \{x^i_j| \forall t(x^i_j) = t(x^i_h),t(y^i_j)=1\}$, $-\Omega^i_l = \{x^i_j| \forall t(x^i_j) = t(x^i_h),t(y^i_j)=-1\}$, depending on trend features of sample labels. And $+\Omega^i_l \cup -\Omega^i_l = \Omega^i_l$, $ +\Omega^i_l \cap -\Omega^i_l = \varnothing$.

Therefore, based on the division of the above subsets, $S(t(y^i_0)=1 | x^i_0)$ and $S(t(y^i_0)=1 | t(x^i_0))$ are calculated using Equations \ref{equation:trend_score2}, Equations \ref{equation:ori_score2} respectively. 

\subsubsection{Introducing the concept of membership degree}

In Equations \ref{equation:trend_score2} and \ref{equation:ori_score2}, $U_{\Omega^i_l}(x^i_0)$ denotes the degree to which $x^i_0$ is subordinate to $\Omega^i_l$, which is the concept of degree of membership function in fuzzy sets. $U_{\Omega^i_l}(x^i_0) \in [0,1]$.

$U_{\Omega^i_l}(x^i_0)$ is introduced to minimize the disturbance caused by random errors. For example, $\Omega^5_1 = \{x_j | t(x_j) = (1,1,-1,1)\}, \Omega^5_2 = \{x_j | t(x_j) = (1,1,1,1)\}, x=(15,16,17,16.9,18)$. Under the interference of random errors, $x$ becomes $\hat{x}=(15,16,17,17.1,18)$. Apparently, $x \in \Omega^5_1$, but $\hat{x} \in \Omega^5_2$. Therefore, the subset to which the sample $x$ belongs changes under the interference of random errors. So, in order to reduce the interference generated by random errors, $U_{\Omega^i_l}(x^i_0)$ is introduced in this paper. 

In this case, $U_{\Omega^5_1}(x)=1,U_{\Omega^5_2}(x)=0.75$ and $U_{\Omega^5_1}(\hat{x})=0.75,U_{\Omega^5_2}(x)=1$. So, even though random interference makes $x$ become $\hat{x}$, the introduction of $U_{\Omega^i_l}(x^i_0)$ makes $\hat{x}$ still belong to a large extent to $\Omega^5_1$, the subset to which the real sample $x$ belongs.

\subsubsection{Probability calculation}

In Equations \ref{equation:trend_score1}, $S(t(y^i_0)=1 | x^i_0 \in \Omega^i_l, t(x^i_0))$ denotes score of $t(y^i_0)=1$ computed based on $t(x^i_0)$ when $x^i_0 \in \Omega^i_l$. Assuming that the sample size of $\Omega^i_l$ is $200$, of which $100$ samples belong to $+\Omega^i_l$, the frequency result of $S(t(y^i_0)=1 | x^i_0 \in \Omega^i_l, t(x^i_0))$ is $0.5$. Assuming that the sample size of $\Omega^i_l$ is $60$, of which $30$ samples belong to $+\Omega^i_l$, the frequency result of $S(t(y^i_0)=1 | x^i_0 \in \Omega^i_l, t(x^i_0))$ is also $0.5$. Nevertheless, we believe that the former result is more reliable because it is based on a larger sample size. 

Therefore, in this paper, we discard the pure frequency result to calculate $S(t(y^i_0)=1 | x^i_0 \in \Omega^i_l, t(x^i_0))$ and use Bayesian estimation to calculate it. Since the conjugate distribution of the binomial distribution is the beta distribution $\beta(a,b)$, in this paper we set the prior distribution to be $\beta(1,1)$. After calculation, the corresponding posterior distribution is $\beta(|+\Omega^i_l|+1,|-\Omega^i_l|+1)$. Therefore, $S(t(y^i_0)=1 | x^i_0 \in \Omega^i_l, t(x^i_0)) = \frac{|+\Omega^i_l|+1}{|+\Omega^i_l|+|-\Omega^i_l|+2}$.

In Equations \ref{equation:ori_score1}, $S(t(y^i_0)=1 | x^i_0 \in \Omega^i_l, x^i_0)$ denotes score of $t(y^i_0)=1$ computed based on $x^i_0$ when $x^i_0 \in \Omega^i_l$.  $S(t(y^i_0)=1 | x^i_0 \in \Omega^i_l, x^i_0)$ is defined as the Mahalanobis distance between $x^i_0$ and $+\Omega^i_l$, and this Mahalanobis distance is mapped to $[0,1]$.

$m$ is a threshold used to determine whether a sample set can be called a large sample set, which is usually equal to $50$. when the sample size in $\Omega^i_l$ is smaller, the error generated by Equations \ref{equation:ori_score1} and \ref{equation:trend_score1} is larger, so, in this paper, we use conditional judgment to select the appropriate lag order $i$ such that all the $\Omega^i_l$ corresponding to lag order $i$ satisfy $|\Omega^i_l|>m$.

Therefore, the above description reveals that TeMoP has the following advantages: 
\begin{itemize}
    \item By integrating various models trained based on samples with different lag orders, TeMoP has excellent robustness.
    \item By introducing trend features, TeMoP has small prediction errors.
\end{itemize}

In the next section \ref{sec:data_results}, an empirical analysis based on real-world data sets will be conducted to validate the above modeling merits.

\section{Experiments}\label{sec:data_results}

This section first provides a detailed description of the experimental data, assessment metrics, comparison models, experimental setup and experimental schemes. Then, the comparison results are fully analyzed.

\subsection{Data description}

  There are nine major global financial market indices as data sets to assess models. In the experiment, the daily closing price of each index is selected for analysis. Table \ref{tab:data_set_information} provides the basic information of the selected stock index data sets. All data sets can be obtained from \href{https://cn.investing.com/}{https://cn.investing.com/}.

  \begin{table}[htbp!]
    \centering
    \footnotesize
    \setlength\tabcolsep{1pt}
    \caption{Statistical summary for data sets}\label{tab:data_set_information}
    \begin{tabular}{lcccccc}
    \toprule
        Data set & Abbreviation & Market type & Std$^a$ & LB(30)$^b$ & ADF$^c$ & JB$^d$ \\
        \midrule
        Dow\_Jones\_Average & DJIA & Developed & 5630.55  & 0.000$^e$  & 0.967$^e$  & 0.000$^e$  \\ 
        FTSE\_China A50 & FCA & Developing & 3790.36  & 0.000  & 0.286  & 0.000  \\ 
        Hang\_Seng & HS & Semi-Developed & 4693.36  & 0.000  & 0.164  & 0.000  \\ 
        NASDAQ\_Composite & NC & Developed & 2235.40  & 0.000  & 0.999  & 0.000  \\ 
        Nifty\_50 & Nif & Developing & 2986.29  & 0.000  & 0.916  & 0.000  \\ 
        Nikkei\_225 & Nik & Semi-Developed &  4809.51  & 0.000  & 0.927  & 0.000  \\ 
        S\&P\_500 & SP & Developed & 653.28  & 0.000  & 0.988  & 0.000  \\ 
        Shanghai\_Composite & SC & Developing & 872.61  & 0.000  & 0.193  & 0.000  \\ 
        SZSE\_Component & SZC & Developing & 3518.29  & 0.000  & 0.170  & 0.000 \\ 
        \bottomrule
    \end{tabular}
    \begin{flushleft}
         $^{a}$ Std denotes the standard deviation of data sets. $^{b}$ LB(30) denotes the Ljung-Box test for a series in the $30$th order lag range. $^{c}$ ADF stands for ADF test (also called unit root test). $^d$ JB stands for Jarque-Bera test.
         $^{e}$ The significance level for all tests is $0.05$.
    \end{flushleft}
\end{table}

  According to the standard deviation, SP and SC exhibit lower volatility, while DJIA, HS, and Nik are the most volatile. Then, the p-value of the LB test rejects the original hypothesis that the data is white noise, indicating that all data sets hold significance for further analysis. Moreover, the p-value of the ADF test rejects the original hypothesis of no unit root in the data, implying that all data sets in our experiment are deemed non-stationary. Finally, the p-value results of the JB test, utilized to assess data adherence to a normal distribution, suggest that none of the data sets conform to a normal distribution.

\begin{figure}[htbp!]
\centering
\includegraphics[scale=0.45]{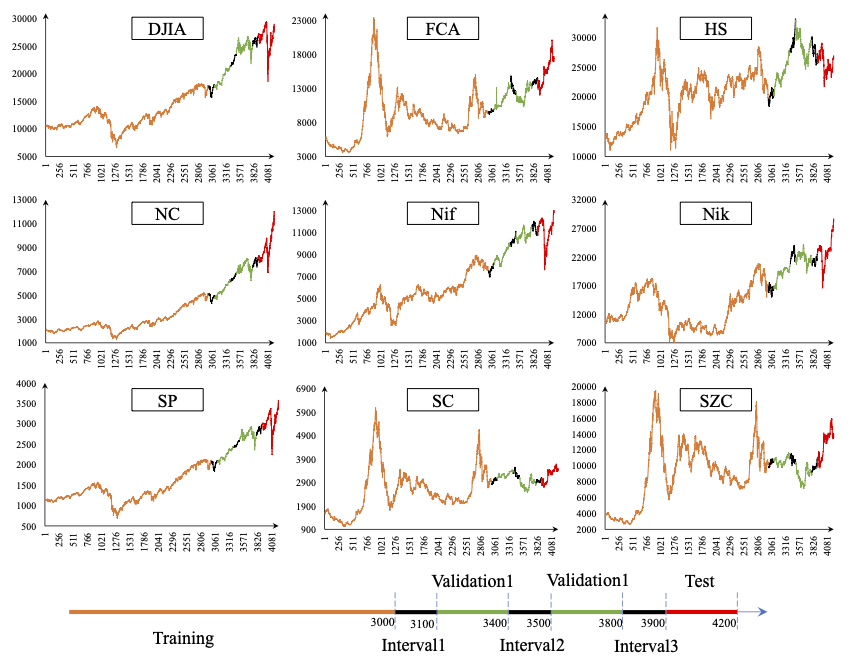}
\caption{Division of data sets. The orange indicates training set. The black indicates time intervals. The green indicates validation set, and the red indicates test set.}\label{fig:data_sets}
\end{figure}

  In order to provide a comprehensive and objective assessment of models, the data sets are divided into four parts: a training set, two validation sets and a test set. The size of the training set is fixed at $3000$. The size of every validation set is fixed at $300$. The test set consists of $300$ data points. Additionally, to prevent information leakage, there are time intervals between different sets. Figure \ref{fig:data_sets} illustrates the division of different data sets.

Time series trend forecasting is often viewed as a binary classification problem. If the closing price of a stock at day $t$ is higher than its closing price at day $t-1$, the stock movement trend is defined as “upward” ($y_t = 1$), otherwise as “downward” ($y_t = -1$). Table \ref{tab:data_set_ratio} shows proportion of positive (“upward”) samples in every data set. The results show that the proportions of positive (“upward”) and negative (“downward”) samples are very close to each other. Hence, the data sets are roughly balanced.

\begin{table}[htbp!]
    \centering
    \footnotesize
    \caption{Proportion of positive (“upward”) samples in data sets}\label{tab:data_set_ratio}
    \begin{tabular}{lcccccc}
    \toprule
        Data & Training & Validation$1$ & Validation$2$ & Test \\
        \midrule
        DJIA  & 53.3\% & 50.5\%	& 55.2\% & 56.5\%  \\ 
        FCA & 51.1\% & 56.9\% &	48.5\% & 54.5\%  \\ 
        HS & 52.2\% & 57.2\% &	53.5\% & 52.8\%  \\ 
        NC & 54.7\%	& 58.2\% & 54.8\% & 59.9\%  \\ 
        Nif & 53.6\% & 54.8\% & 53.5\% & 56.5\%  \\ 
        Nik &  52.0\% & 52.5\% & 53.8\%	& 51.2\%  \\ 
        SP & 54.8\% & 52.2\% & 53.8\% & 58.9\%  \\ 
        SC & 53.4\% & 57.2\% & 47.5\% & 53.2\%  \\ 
        SZC & 52.1\% & 52.5\% & 44.8\% & 56.2\% \\ 
        \bottomrule
    \end{tabular}
\end{table}

\subsection{Evaluation metrics}

In order to comprehensively assess prediction results, We adopt Accuracy (ACC), the area under the precision–recall curve (AUC), F$1$-score (F$1$) and Sharpe Ratio (SR), each of which measures a different aspect of the results. 

\textbf{Accuracy (ACC):} ACC is used to calculate the percentage of samples that are correctly predicted out of all samples with predictions. The mathematical expression for ACC is shown in Equation \ref{equation:acc}, where TP is the true positive, TN is the true negative, FP is the false positive and FN is the false negative.
\begin{equation}\label{equation:acc}
    ACC = \frac{TP+TN}{TP+FP+TN+FN}
\end{equation}

\textbf{F$1$-score (F$1$):} In contrast to ACC, F$1$ measures the performance of the model in positive and negative samples, respectively. F$1$ can evaluate the prediction performance of the model more comprehensively. The mathematical expression for F$1$ is shown in Equation \ref{equation:f1}.
\begin{equation}\label{equation:f1}
    F1 = \frac{2 \times PRE \times REC}{PRE + REC},
\end{equation}
where
\begin{align*}
    PRE = \frac{TP}{TP+FP},\\
    REC = \frac{TP}{TP+FN}.\\
\end{align*}

\textbf{Sharpe Ratio (SR):} SR is used to evaluate the investment performance of models in the real stock market. The larger the value, the higher the profitability of the model's forecast results. The mathematical expression for SR is shown in Equation \ref{equation:sr}. 

\begin{equation}\label{equation:sr}
    SR = \frac{E(R_p) - R_f}{\sigma_p}
\end{equation}

\subsection{Models for comparison}

The experiment selects several models from the fields of statistics, machine learning in narrow sense, and deep learning respectively for comparative analysis with TeMoP. Here is a short description of these models.

\begin{itemize}
      \item \textbf{Logistic Regression (LR):} LR is a classical model used in statistical learning for classification tasks. The model has good interpretability.
      \item \textbf{Support Vector Machines(SVM):} SVM is a novel small-sample learning method with a solid theoretical foundation, strong generalization ability and easy interpretation of results.
      \item \textbf{Random Forests (RF):} RF is a classical model in the field of machine learning. Based on the bagging integration algorithm, RF has good robustness.
      \item \textbf{Light Gradient Boosting Machine (LGBM):} LGBM is also a classic model in the field of machine learning, based on the Boosting integration algorithm, LGBM has better prediction accuracy.
      \item \textbf{Convolutional Neural Network (CNN):} CNN, a foundational deep learning model, leverages convolution and pooling techniques to extract intricate features from input data efficiently. 
      \item \textbf{Long Short term Memory (LSTM):} Unlike traditional feedforward neural networks, LSTM is recurrent, enabling it to capture long-term dependencies in sequence data.
      \item \textbf{ResCNN:} ResCNN \citep{zou2019integration}, a state-of-the-art model, integrates residual network with convolutional neural network (CNN), for time series classification. 
      \item \textbf{InceptionTime:} The emergence of InceptionTime \citep{ismail2020inceptiontime} marks an important milestone in the field of time series classification. It not only reaches the level of accuracy of previous state-of-the-art methods, but also achieves a qualitative leap in scalability. 
  \end{itemize}

\subsection{Experimental settings}

The experiment is based on windows10 system. The programming language is Python3.8.1, and the corresponding compiler is PyCharm2020.1.2. The main Python libraries used in the experiment are numpy1.23.3, pandas1.4.4, tensorflow2.10.0, keras2.10.0, scikit-learn1.1.3, and tsai0.3.9.

\subsection{Experimental schemes}

Compared to the comparison models, TeMoP is not only a non-parametric model, but also can adaptively choose the maximum lag order according to the nature of the data. Therefore, TeMoP only needs to be trained on Training and prediction performance is evaluated on Test based on different metrics. Therefore, for the fairness of the comparison between different models, two schemes are designed as follows. The hyper-parameter for comparison models is shown in Table \ref{tab:models_hyper-parameter}.

\begin{itemize}
      \item \textbf{Scheme $1$:} For each data set, the lag order of samples used to train comparison models is consistent with the maximum lag order of TeMoP. Then, comparison models are trained on Training, and a set of optimal hyper-parameters  is selected on Validation$1$ using the indicator F$1$ by grid search method. Finally, the prediction performance is evaluated on Test using different metrics. 
      \item \textbf{Scheme $2$:} Let the lag order $i \in [3,30]$. For each data set, each comparison model is trained on Training, and a set of optimal hyper-parameters is selected on Validation$1$ using metric F$1$ by grid search method, and the optimal lag order is selected on Validation$2$ using metric F$1$. Finally, the predictive performance is evaluated on Test using different metrics. Since a large amount of literature \citep{zhao2022stock,dudek2023std} suggests that the number of lag orders generally does not exceed $30$, and furthermore, since CNN is one of the comparison models and the hyper-parameter kernel\_size in CNN generally takes a value of not less than $3$, let the lag order $i \in [3,30]$. 
  \end{itemize}

The difference between Scheme $1$ and $2$ is that the former assigns lag
orders to the comparison models, and the latter selects the optimal lag order
based on the performance of different lag orders of the comparison models
on the data sets.

\begin{table}[htbp!]
    \centering
    \footnotesize
    \caption{Hyper-parameter for comparison models}\label{tab:models_hyper-parameter}
    \begin{tabular}{lcccccc}
    \toprule
        Model & Library & Hyper-paramete & Value \\
        \midrule
        LR & Scikit-learn & None & None \\
        SVM & Scikit-learn & C & [0.1,0.5,1] \\
        \multirow{3}{*}{RF} & \multirow{3}{*}{Scikit-learn} & n\_estimators & [50,100,150] \\
        & & max\_depth & [5,7,9] \\
        & & min\_samples\_split & [5,10] \\
        \multirow{3}{*}{LGBM} & \multirow{3}{*}{Scikit-learn} & learning\_rate & [0.005,0.01,0.02]\\
        & & min\_child\_samples & [10,30,50] \\
        & & max\_depth & [4,5,6] \\
        \multirow{3}{*}{CNN} & \multirow{3}{*}{Keras} & filters & [16,32,64] \\
        & & batch\_size & [64,128,256]\\
        & & epochs & [20,50]\\
        \multirow{3}{*}{LSTM} & \multirow{3}{*}{Keras} & units & [16,32,64] \\
        & & batch\_size & [64,128,256] \\
        & & epochs & [20,50]\\
        \multirow{2}{*}{ResCNN} & \multirow{2}{*}{Tsai} & epochs & [15,30] \\
        & & learning\_rate & [0.005,0.01,0.02] \\
        \multirow{2}{*}{InceptionTime} & \multirow{2}{*}{Tsai} & epochs & $[15,30]$ \\
        & & learning\_rate & [0.005,0.01,0.02] \\
        \bottomrule
    \end{tabular}
\end{table}

\subsection{Experimental results}


\subsubsection{Scheme $1$}

\begin{table}[htbp!]
    \centering
    \footnotesize
    \setlength\tabcolsep{1pt}
    \caption{Optimal hyper-parameters of models for scheme $1$}\label{tab:models_hyper-parameter_1}
    \begin{tabular}{lcccccccccc}
    \toprule
        Model & Hyper-paramete & DJIA & FCA & HS & NC & Nif & Nik & SP & SC & SZC\\
        \midrule
        SVM & C & 0.1 & 0.1 & 0.1 & 0.1 & 0.1 & 0.1 & 0.1 & 0.1 & 0.1\\
        \multirow{3}{*}{RF} & n\_estimators & 150 & 150 & 100 & 50 & 100 & 100 & 50 & 50 & 100\\
        & max\_depth & 7 & 5 & 9 & 5 & 9 & 7 & 7 & 5 & 5\\
        & min\_samples\_split & 10 & 5 & 10 & 10 & 5 & 10 & 5 & 5 & 5 \\
        \multirow{3}{*}{LGBM} & learning\_rate & 0.005 & 0.005 & 0.005 & 0.005 & 0.005 & 0.005 & 0.005 & 0.005 & 0.005 \\
        & min\_child\_samples & 10 & 10 & 10 & 10 & 10 & 10 & 10 & 10 & 10 \\
        & max\_depth & 4 & 4 & 4 & 4 & 4 & 4 & 4 & 4 & 4 \\
        \multirow{3}{*}{CNN} & filters & 64 &	16 & 64 & 64 & 32	& 64	& 16	& 32	& 32 \\
        & batch\_size & 256 & 256	& 256	& 64 & 256	& 256	& 64	& 256	& 128 \\
        & epochs & 20	& 20	& 50	& 50	& 50	& 20	& 20	& 20	& 50\\
        \multirow{3}{*}{LSTM} & units & 16 & 32	& 16	& 64	& 64	& 16	& 16	& 16	& 64 \\
        & batch\_size & 64 & 128	& 64	& 128	& 64	& 128	& 64	& 64	& 64 \\
        & epochs & 20	& 50	& 20	& 20	& 50	& 50	& 20	& 20	& 50\\
        \multirow{2}{*}{ResCNN} & epochs & 30	& 15	& 30	& 30	& 15	& 15	& 15	& 15	& 15 \\
        & learning\_rate & 0.02	& 0.005	& 0.02	& 0.005 &	0.02	& 0.005	& 0.01	& 0.005	& 0.01 \\
        \multirow{2}{*}{InceptionTime} & epochs & 15	& 15	& 15	& 15	& 15	& 15	& 15	& 15	& 15 \\
        & learning\_rate & 0.01	& 0.01	& 0.01	& 0.01	& 0.02	& 0.005	& 0.01	& 0.005	& 0.005 \\
        \bottomrule
    \end{tabular}
\end{table}

After training TeMoP on Training for each data set, the maximum lag order of TeMoP on data set SP is $5$, and the maximum lag order on the other data sets is $6$. Therefore, let the lag order of comparison models and the maximum lag order of TeMoP be the same on each data set. Then, each comparison model is trained on Training, and a set of optimal hyper-parameters is selected on Validation$1$ using the indicator F$1$ by grid search method. Table \ref{tab:models_hyper-parameter_1} shows that optimal hyper-parameters for each model on different data sets. 

\begin{figure}[htbp!]
\centering
\includegraphics[scale=0.32]{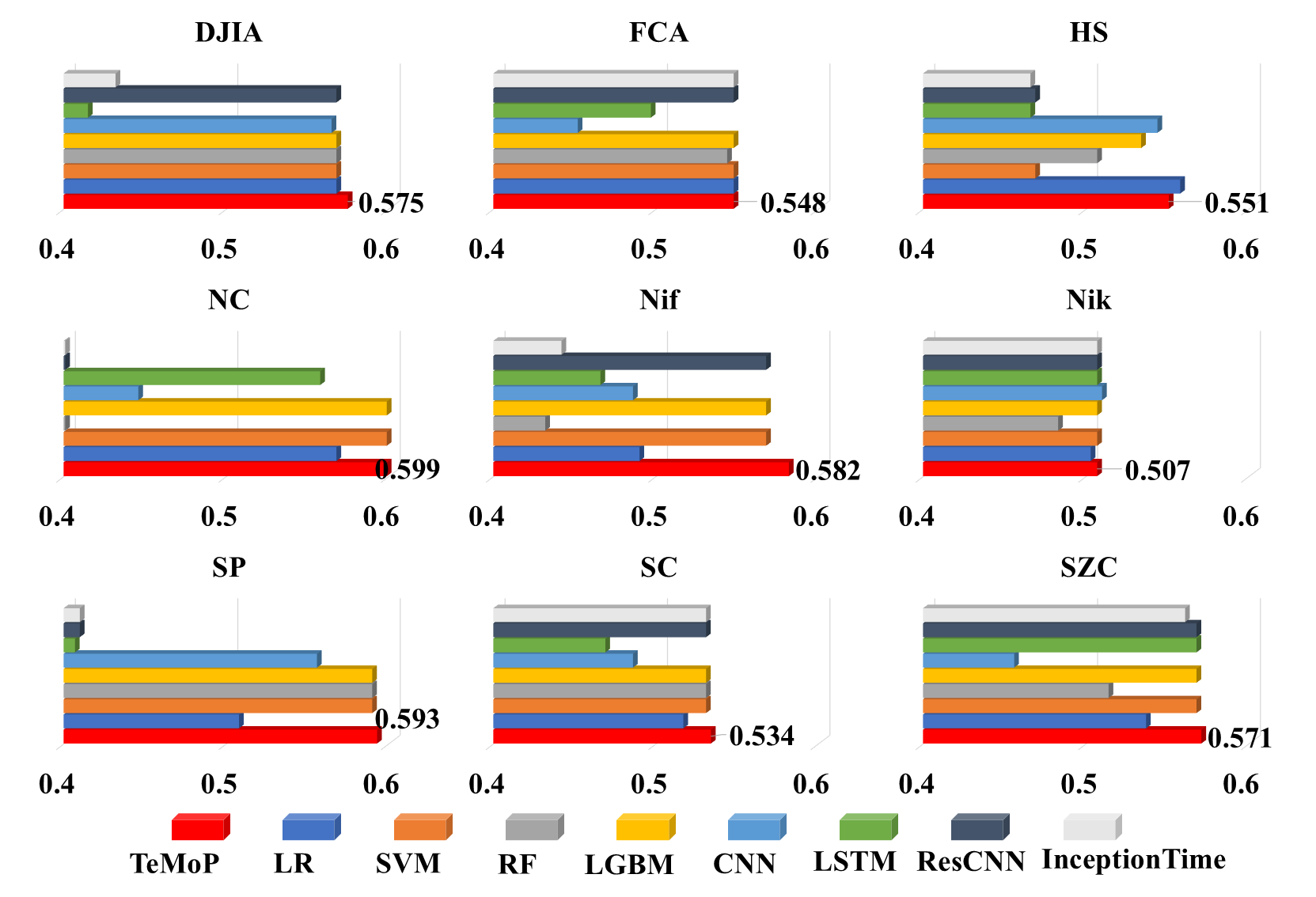}
\caption{ACC of models on different data sets for scheme $1$}\label{fig:scheme1_acc}
\end{figure}

\begin{figure}[htbp!]
\centering
\includegraphics[scale=0.32]{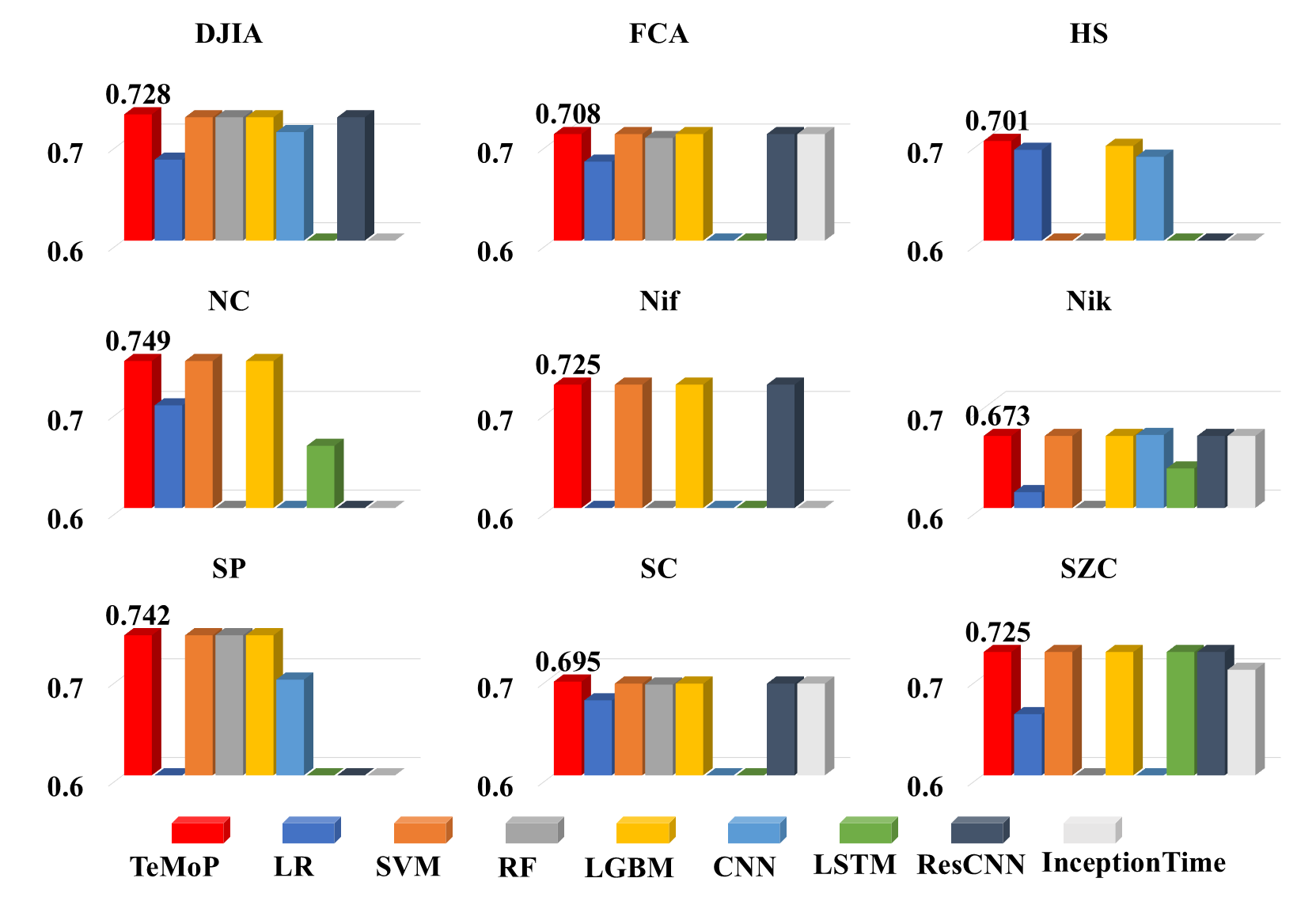}
\caption{F$1$ of models on different data sets for scheme $1$}\label{fig:scheme1_f1}
\end{figure}

\begin{figure}[htbp!]
\centering
\includegraphics[scale=0.32]{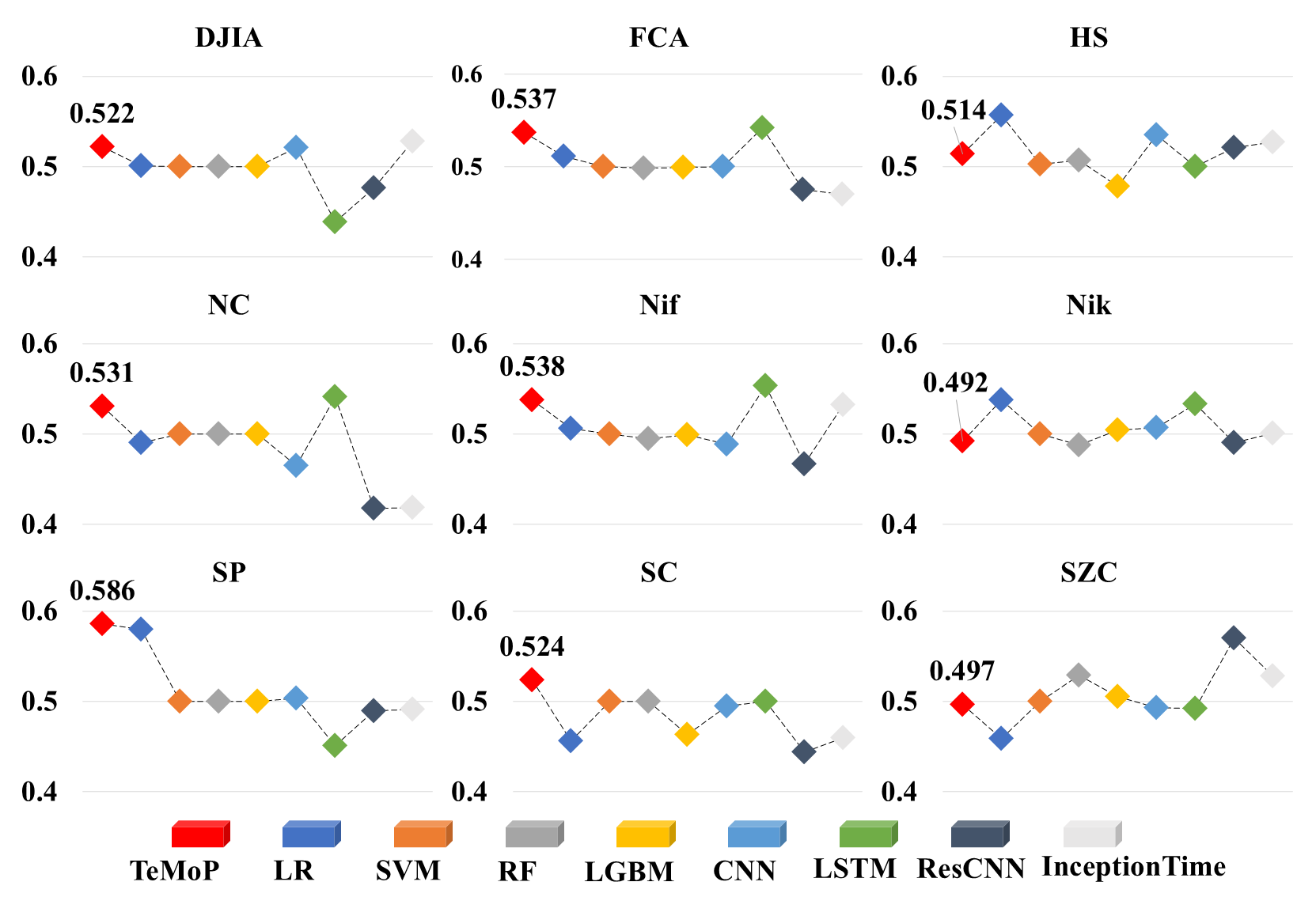}
\caption{AUC of models on different data sets for scheme $1$}\label{fig:scheme1_auc}
\end{figure}

\begin{figure}[htbp!]
\centering
\includegraphics[scale=0.32]{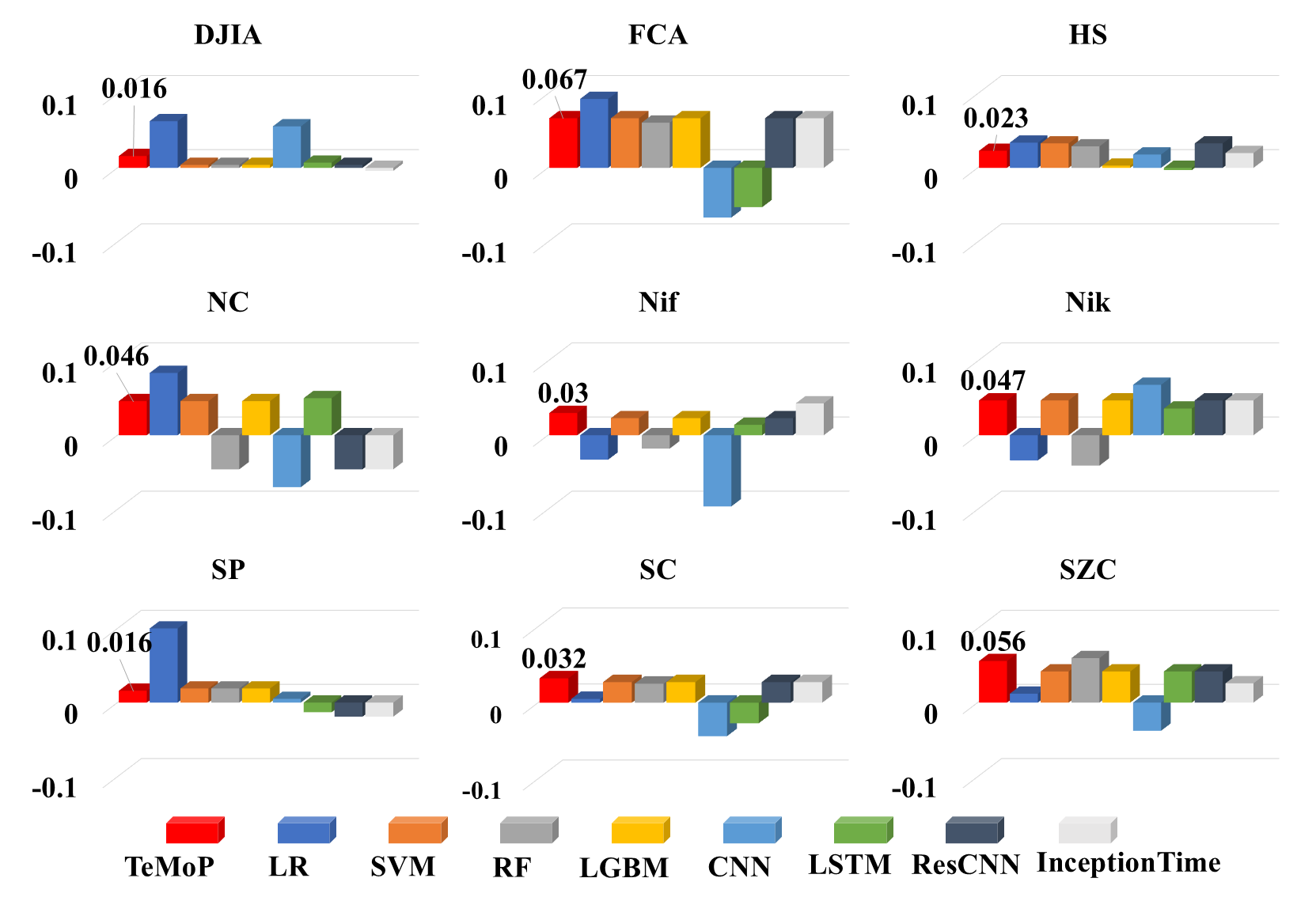}
\caption{SR of models on different data sets for scheme $1$}\label{fig:scheme1_sr}
\end{figure}

Finally, the prediction performance of all models is evaluated on Test using different metrics. Figure \ref{fig:scheme1_acc}, \ref{fig:scheme1_f1}, \ref{fig:scheme1_auc}, \ref{fig:scheme1_sr} show ACC, AUC, F$1$, SR of models on different data sets for scheme $1$ respectively. Specific numerical values are detailed in Table \ref{tab:appendix_acc1}, \ref{tab:appendix_f11}, \ref{tab:appendix_auc1}, \ref{tab:appendix_sr1}. Table \ref{tab:all_metrics1} shows the combined prediction performance of models on all data sets.

\begin{table}[htbp!]
    \centering
    \footnotesize
    \setlength\tabcolsep{1pt}
    \caption{ACC of models on different data sets for scheme $1$}\label{tab:appendix_acc1}
    \begin{tabular}{lccccccccccc}
    \toprule
        model & DJIA & FCA & HS & NC & Nif & Nik & SP & SC & SZC & mean & std \\ 
    \midrule
        TeMoP & 0.575 &	0.548 &	0.551 &	0.599 &	0.582 &	0.507 &	0.593 &	0.534 &	0.571 & 0.562 & 0.030  \\
        LR & 0.568 & 0.548 &	0.558 &	0.568 &	0.490 &	0.503 &	0.508 &	0.517 &	0.537 & 0.533 & 0.029  \\
        SVM & 0.568 &	0.548 &	0.469 &	0.599 &	0.568 &	0.507 &	0.590 &	0.531 &	0.568  & 0.550  & 0.041  \\
        RF & 0.568 &	0.544 &	0.507 &	0.401 &	0.432 &	0.483 &	0.590 &	0.531 &	0.514 & 0.508  & 0.061  \\
        LGBM & 0.568 &	0.548 &	0.534 &	0.599 & 0.568 &	0.507 &	0.590 &	0.531 &	0.568  & 0.557  & 0.030  \\
        CNN & 0.565 &	0.452 &	0.544 &	0.446 &	0.486 &	0.510 &	0.556 &	0.486 &	0.456  & 0.500  & 0.046  \\
        LSTM & 0.415 &	0.497 &	0.466 &	0.558 &	0.466 &	0.507 &	0.407 &	0.469 &	0.568  & 0.483  & 0.055  \\
        ResCNN & 0.568 	& 0.548 	& 0.469 	& 0.401 	& 0.568	& 0.507 	& 0.410 & 0.531 	& 0.568  & 0.508  & 0.066  \\
        InceptionTime & 0.432 &	0.548 &	0.466 	& 0.401 &	0.442 &	0.507 &	0.410 &	0.531 	& 0.561  & 0.477  & 0.660 \\
    \bottomrule
    \end{tabular}
\end{table}

In Figure \ref{fig:scheme1_acc} and Table \ref{tab:appendix_acc1}, in addition to the data sets HS, Nik, TeMoP has the largest ACC on the remaining data sets. For the data set HS, compared to the largest ACC $0.558$, ACC of TeMoP is $0.551$, with a $1.2\%$ difference. For data set Nik, compared to the largest ACC $0.510$, ACC of TeMoP is $0.507$, a difference of $0.6\%$. Therefore, for the data sets HS,Nik, the ACC of TeMoP is very close to the maximum ACC. This indicates that TeMoP performs very well on the metric ACC for each data set. In addition, among all the models, the ACC of just two models TeMoP and LGBM are greater than $0.5$ on each data set, which indicates that TeMoP and LGBM have good robustness across data sets compared to the rest of the models. The mean and standard deviation of the ACC results of all models on different data sets are shown in Table \ref{tab:all_metrics1}, which are used to measure the robustness of the models between different data sets. In Table \ref{tab:all_metrics1}, the mean ACC of TeMoP takes the largest value among all models. The standard deviation of ACC of TeMoP, LGBM, and LR is smaller compared to other models. Therefore, TeMoP exhibits the smallest prediction error and good robustness across data sets when measuring the prediction performance of the models based on the metric ACC.

\begin{table}[htbp!]
    \centering
    \footnotesize
    \setlength\tabcolsep{1pt}
    \caption{F$1$ of models on different data sets for scheme $1$}\label{tab:appendix_f11}
    \begin{tabular}{lccccccccccc}
    \toprule
        model & DJIA & FCA & HS & NC & Nif & Nik & SP & SC & SZC & mean & std \\
    \midrule
        TeMoP & 0.728 &	0.708 &	0.701 &	0.749 &	0.725 &	0.673 	& 0.742 &	0.695 &	0.725 &	0.716 &	0.024  \\
        LR & 0.682  & 0.680  & 0.692  & 0.704  & 0.380  & 0.616  & 0.457  & 0.676  & 0.662  & 0.616  & 0.110  \\
        SVM & 0.725  & 0.708  & 0.025  & 0.749  & 0.725  & 0.673  & 0.742  & 0.693  & 0.725  & 0.640  & 0.219  \\ 
        RF & 0.725  & 0.704  & 0.525  & 0.000  & 0.056  & 0.232  & 0.742  & 0.692  & 0.491  & 0.463  & 0.278  \\ 
        LGBM & 0.725  & 0.708  & 0.696  & 0.749  & 0.725  & 0.673  & 0.742  & 0.693  & 0.725  & 0.715  & 0.023  \\ 
        CNN & 0.710  & 0.000  & 0.685  & 0.318  & 0.470  & 0.674  & 0.697  & 0.575  & 0.216  & 0.483  & 0.239  \\ 
        LSTM & 0.196  & 0.308  & 0.000  & 0.663  & 0.219  & 0.640  & 0.163  & 0.000  & 0.725  & 0.324  & 0.266  \\ 
        ResCNN & 0.725  & 0.708  & 0.037  & 0.000  & 0.725  & 0.673  & 0.000  & 0.693  & 0.725  & 0.476  & 0.328  \\ 
        InceptionTime  & 0.000  & 0.708  & 0.037  & 0.000  & 0.079  & 0.673  & 0.000  & 0.693  & 0.707  & 0.322  & 0.335 \\
    \bottomrule
    \end{tabular}
\end{table}

Unlike ACC, F$1$ allows a more objective assessment of the models' prediction accuracy. In Figure \ref{fig:scheme1_f1} and Table \ref{tab:appendix_f11}, TeMoP has the largest F$1$ on the rest of the data sets except for data set Nik. For data set Nik, compared to the largest F$1$ $0.674$, F$1$ of TeMoP is $0.637$, a difference of $0.1\%$, which is very close to the largest F$1$. This indicates that the F$1$ of TeMoP is the largest in almost every data set. The consistent results between different data sets also reflect that QQ has good robustness. The mean and standard deviation of the F$1$ results of all models on different data sets are shown in Table \ref{tab:all_metrics1}. In Table \ref{tab:all_metrics1}, the mean of F$1$ for TeMoP takes the largest value among all models. The standard deviation of F$1$ of TeMoP is smaller compared to other models. Therefore, TeMoP exhibits minimal prediction error and good robustness across different data sets when measuring the prediction performance of the model based on the metric F$1$.

\begin{table}[htbp!]
    \centering
    \footnotesize
    \setlength\tabcolsep{1pt}
    \caption{AUC of models on different data sets for scheme $1$}\label{tab:appendix_auc1}
    \begin{tabular}{lccccccccccc}
    \toprule
        model & DJIA & FCA & HS & NC & Nif & Nik & SP & SC & SZC & mean & std \\ 
    \midrule
        TeMoP & 0.522 &	0.537 &	0.514 &	0.531 &	0.538 &	0.492 &	0.586 &	0.524 &	0.497  & 0.527 & 0.027  \\
        LR & 0.501 &	0.511 &	0.557 &	0.490 	& 0.506 &	0.538 &	0.580 &	0.456 &	0.459  & 0.511 & 0.041  \\
        SVM & 0.500 &	0.500 &	0.503 &	0.500 	& 0.500 &	0.500 &	0.500 &	0.500 &	0.500  & 0.500  & 0.001  \\
        RF & 0.500 &	0.498 &	0.507 &	0.500 	& 0.495 &	0.488 &	0.500 &	0.500 &	0.529  & 0.502  & 0.011  \\
        LGBM & 0.500 &	0.499 &	0.478 &	0.500 	& 0.499 &	0.504 &	0.500 &	0.463 &	0.505  & 0.494  & 0.014  \\
        CNN & 0.521 &	0.500 &	0.535 &	0.465 	& 0.489 &	0.507 &	0.504 &	0.495 &	0.493  & 0.501  & 0.020  \\
        LSTM & 0.439 &	0.542 &	0.500 &	0.541 &	0.553 &	0.533 &	0.451 &	0.500 &	0.492  & 0.505  & 0.041  \\
        ResCNN & 0.476 &	0.475 &	0.521 &	0.418 &	0.467 &	0.490 &	0.490 &	0.444 &	0.570  & 0.483  & 0.043  \\
        InceptionTime & 0.528 &	0.470 &	0.527 &	0.419 &	0.532 &	0.501 &	0.491 &	0.460 &	0.528  & 0.495  & 0.039  \\
    \bottomrule
    \end{tabular}
\end{table}

AUC measures the model's ranking ability. In Figure \ref{fig:scheme1_auc} and Table \ref{tab:appendix_auc1}, AUC of TeMoP is more than $0.5$ on all data sets except data sets Nik and SZC. TeMoP has the largest AUC in data sets SP and SC. TeMoP has the second largest AUC in data sets DJIA, FCA, NC, and Nif. The mean and standard deviation of the AUC results of all the models on different data sets are shown in Table \ref{tab:all_metrics1}. Table \ref{tab:all_metrics1} shows that TeMoP has the largest mean AUC among all models, and the standard deviation AUC for TeMoP is smaller compared to the other models. Compared to the second largest mean AUC, mean AUC of TeMoP is $4$ percentage points higher. Compared to the smallest average AUC, mean AUC improves by $10$ percentage points. Therefore, TeMoP shows minimal prediction error and good robustness across data sets when measuring the predictive performance of the model based on the metric AUC.

\begin{table}[htbp!]
    \centering
    \footnotesize
    \setlength\tabcolsep{1pt}
    \caption{SR of models on different data sets for scheme $1$}\label{tab:appendix_sr1}
    \begin{tabular}{lccccccccccc}
    \toprule
        model & DJIA & FCA & HS & NC & Nif & Nik & SP & SC & SZC & mean & std \\
    \midrule
        TeMoP & 0.016 &	0.067 &	0.023 &	0.046 &	0.030 &	0.047 &	0.016 &	0.032 &	0.056 &	0.037 &	0.018  \\
        LR & 0.063  & 0.093  & 0.034  & 0.084  & -0.033  & -0.034  & 0.131  & 0.005  & 0.012  & 0.040  & 0.057  \\
        SVM & 0.004  & 0.067  & 0.033  & 0.046  & 0.023  & 0.047  & 0.019  & 0.027  & 0.042  & 0.034  & 0.019  \\
        RF & 0.004  & 0.061  & 0.029  & -0.046  & -0.018  & -0.041  & 0.019  & 0.025  & 0.060  & 0.010  & 0.039  \\
        LGBM & 0.004  & 0.067  & 0.003  & 0.046  & 0.023  & 0.047  & 0.019  & 0.027  & 0.042  & 0.031  & 0.021  \\
        CNN & 0.056  & -0.067  & 0.018  & -0.070  & -0.096  & 0.068  & 0.005  & -0.044  & -0.038  & -0.019  & 0.058  \\ 
        LSTM & 0.007  & -0.053  & -0.003  & 0.050  & 0.014  & 0.036  & -0.013  & -0.027  & 0.042  & 0.006  & 0.034  \\ 
        ResCNN & 0.004  & 0.067  & 0.033  & -0.046  & 0.023  & 0.047  & -0.019  & 0.027  & 0.042  & 0.020  & 0.035  \\
        InceptionTime  & -0.004  & 0.067  & 0.020  & -0.046  & 0.043  & 0.047  & -0.019  & 0.027  & 0.026  & 0.018  & 0.035 \\ 
    \bottomrule
    \end{tabular}
\end{table}

SR responds to the returns obtained by the portfolios based on the models' prediction results. In Figure \ref{fig:scheme1_sr} and Table \ref{tab:appendix_sr1}, only TeMoP, SVM and LGBM have positive values on each data set, which indicates that the portfolios formed based on the prediction results of these models benefit on each stock index, reflecting the superiority and robustness of these models in simulating the returns. The mean and standard deviation of SR for all models on different data sets are shown in Table \ref{tab:all_metrics1}. In Table \ref{tab:all_metrics1}, although the mean SR of LR takes the largest value among all models, the standard deviation SR of LR is larger. In addition, LR has negative simulation returns on the data sets Nif and Nik. Therefore, taken together, TeMoP not only has a more superior mean SR, but also has a smaller standard deviation SR compared to other models. Therefore, when measuring the models' prediction performance based on the metric SR, TeMoP exhibits minimal prediction error and good robustness across different data sets.

\begin{table}[htbp!]
    \centering
    \footnotesize
    \caption{Prediction performance of models for scheme $1$}\label{tab:all_metrics1}
    \begin{tabular}{lcccc}
    \toprule
        model & ACC & F$1$ & AUC & SR \\ 
    \midrule
        TeMoP & \textbf{0.562$\pm$0.030}$^{a}$ & \textbf{0.716$\pm$0.024} & \textbf{0.527$\pm$0.027} & \textbf{0.037$\pm$0.018}\\
        LR & 0.533$\pm$0.029 & 0.616$\pm$0.110 & 0.511$\pm$0.041 & 0.040$\pm$0.057 \\
        SVM & 0.550$\pm$0.041 & 0.640$\pm$0.219 & 0.500$\pm$0.001 & 0.034$\pm$0.019\\
        RF & 0.508$\pm$0.061 & 0.463$\pm$0.278 & 0.502$\pm$0.011 & 0.010$\pm$0.039\\
        LGBM & 0.557$\pm$0.030 & 0.715$\pm$0.023 & 0.494$\pm$0.014 & 0.031$\pm$0.021\\
        CNN & 0.500$\pm$0.046 & 0.483$\pm$0.239 & 0.501$\pm$0.020 & -0.019$\pm$0.058\\
        LSTM & 0.483$\pm$0.055 & 0.324$\pm$0.266 & 0.505$\pm$0.041 & 0.006$\pm$0.034 \\
        ResCNN & 0.508$\pm$0.066 & 0.476$\pm$0.328 & 0.483$\pm$0.043 & 0.020$\pm$0.035\\
        InceptionTime & 0.477$\pm$0.660 & 0.322$\pm$0.335 & 0.495$\pm$0.039 & 0.018$\pm$0.035\\
    \bottomrule
    \end{tabular}
    \begin{flushleft}
         $^{a}$ 0.562$\pm$0.030 denotes mean of this metric $\pm$ standard deviation of this metric between different data sets. $^{b}$ Bold indicates that this result is optimal for all models.
    \end{flushleft}
\end{table}

Combining the performance of the models on different metrics, it can be found that TeMoP shows superiority and robustness in terms of prediction accuracy, sequencing ability, and simulated returns compared to the comparison models. However, the comparative models are less robust, and the prediction accuracy, sorting ability, and simulated returns fluctuate greatly across different data sets. As a result, TeMoP has a better prediction performance compared to the comparison models.

\subsubsection{Scheme $2$}

\begin{table}[htbp!]
    \centering
    \footnotesize
    \setlength\tabcolsep{1pt}
    \caption{Optimal hyper-parameters of comparison models for scheme $2$}\label{tab:models_hyper-parameter_2}
    \begin{tabular}{lcccccccccc}
    \toprule
        Model & Hyper-parameter & DJIA & FCA & HS & NC & Nif & Nik & SP & SC & SZC\\
        \midrule
        SVM & C & 0.1 & 0.1 & 1 & 0.1 & 0.1 & 0.1 & 0.1 & 0.1 & 0.1\\
        \multirow{3}{*}{RF} & n\_estimators & 50 & 100	& 50 & 50 & 150	& 150	& 150	& 50	& 100\\
        & max\_depth & 9 & 5	& 7	& 5	& 9	& 5	& 7	& 5	& 5\\
        & min\_samples\_split & 10 & 10 & 10 & 10 & 10 & 5 & 10 & 5 & 10 \\
        \multirow{3}{*}{LGBM} & learning\_rate & 0.005 & 0.005 & 0.005 & 0.005 & 0.005 & 0.005 & 0.005 & 0.005 & 0.005 \\
        & min\_child\_samples & 10 & 10 & 10 & 10 & 10 & 10 & 10 & 10 & 10 \\
        & max\_depth & 4 & 4 & 4 & 4 & 4 & 4 & 4 & 4 & 4 \\
        \multirow{3}{*}{CNN} & filters & 64	& 32	& 32	& 64	& 32	& 64	& 64	& 64	& 32 \\
        & batch\_size & 64	& 64	& 64	& 256	& 128	& 64	& 128	& 64	& 128 \\
        & epochs & 20	& 20	& 50	& 20	& 20	& 20	& 50	& 20	& 20\\
        \multirow{3}{*}{LSTM} & units & 32	& 16	& 32	& 16	& 32	& 16	& 16	& 64	& 16\\
        & batch\_size & 256	& 256	& 128	& 256	& 128	& 256	& 256	& 128	& 256 \\
        & epochs & 20	& 20	& 50	& 20	& 20	& 20	& 50	& 20	& 20\\
        \multirow{2}{*}{ResCNN} & epochs & 30	& 15	& 15	& 15	& 15	& 15	& 30	& 15	& 30\\
        & learning\_rate & 0.005 &	0.005	& 0.02	& 0.005	& 0.02	& 0.005	& 0.01	& 0.005	& 0.01\\
        \multirow{2}{*}{InceptionTime} & epochs & 15	& 15	& 15	& 30	& 15	& 15	& 15	& 15	& 15 \\
        & learning\_rate & 0.01	& 0.01	& 0.01	& 0.02	& 0.02	& 0.005	& 0.005	& 0.005	& 0.02\\
        \bottomrule
    \end{tabular}
\end{table}

\begin{table}[htbp!]
    \centering
    \footnotesize
    \setlength\tabcolsep{1pt}
    \caption{Optimal lag orders of comparison models for scheme $2$}\label{tab:lag_order}
    \begin{tabular}{lccccccccc}
    \toprule
        model & DJIA & FCA & HS & NC & Nif & Nik & SP & SC & SZC\\
    \midrule
        LR & 9  & 4  & 8  & 7  & 8  & 8  & 7  & 4  & 3 \\ 
        SVM & 8  & 19  & 3  & 7  & 30  & 4  & 24  & 27  & 28 \\ 
        RF & 8  & 15  & 4  & 21  & 30  & 17  & 21  & 27  & 16 \\ 
        LGBM & 8  & 19  & 25  & 7  & 30  & 4  & 21  & 27  & 28 \\ 
        CNN & 12  & 6  & 15  & 3  & 30  & 5  & 19  & 5  & 16 \\ 
        LSTM & 8  & 30  & 6  & 7  & 30  & 22  & 20  & 21  & 30  \\ 
        ResCNN & 8  & 19  & 26  & 21  & 26  & 4  & 9  & 24  & 28  \\ 
        InceptionTime  & 16  & 19  & 14  & 21  & 25  & 4  & 23  & 27  & 23 \\
    \bottomrule
    \end{tabular}
\end{table}

\begin{figure}[htbp!]
\centering
\includegraphics[scale=0.36]{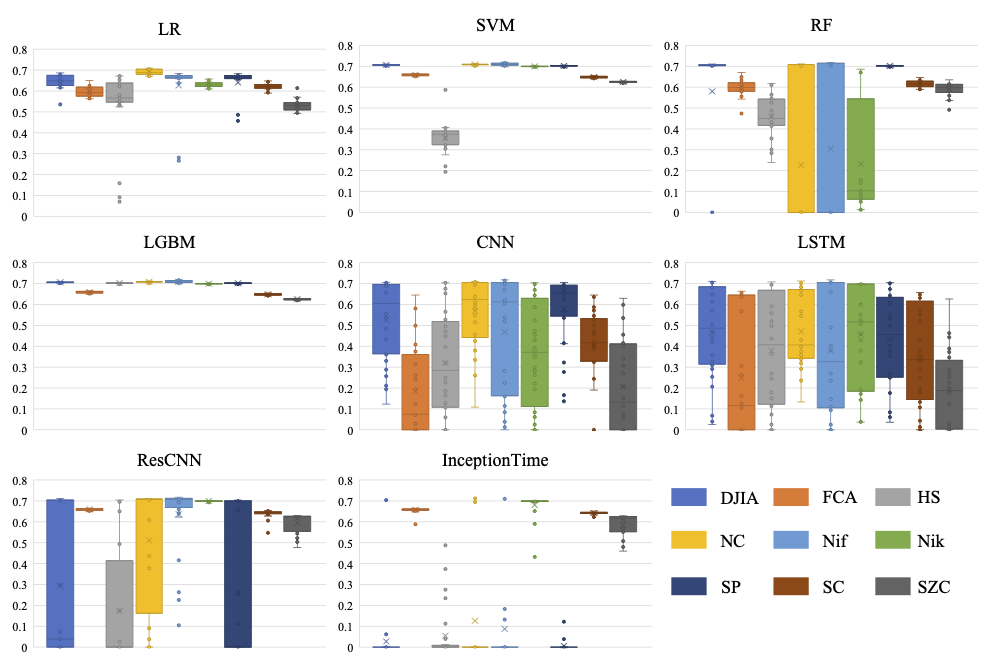}
\caption{F$1$ computed on Validation$2$}\label{fig:robust}
\end{figure}

Let the lag order $i \in [3,30]$. For each data set, each comparison model is trained on Training, and a set of optimal hyper-parameters is selected on Validation$1$ using metric F$1$ by grid search method, and the optimal lag orders are selected on Validation$2$ using metric F$1$. Table \ref{tab:models_hyper-parameter_2} shows that optimal hyper-parameters of the comparison models. Table \ref{tab:lag_order} shows that optimal lag orders of the comparison models. Figure \ref{fig:robust} shows F$1$ computed on Validation$2$. 

First, RF, CNN, LSTM, ResCNN and InceptionTime all have F$1$ equal to $0$ on certain data sets and certain lag orders, which suggests that the predictive ability of these models fails. Then, the predictive performance of these models varies greatly between data sets. In addition, the predictive performance of these models varies greatly between different lag orders on the same data set. This suggests that the robustness of these models is poor and is greatly affected by different data sets and different lag orders. Among all the compared models, LGBM has the optimal robustness. In Figure \ref{fig:robust}, the F$1$ of LGBM based on different lag orders are very concentrated in the same data set. The F$1$ of LGBM are also more stable between different data sets.

\begin{figure}[htbp!]
\centering
\includegraphics[scale=0.32]{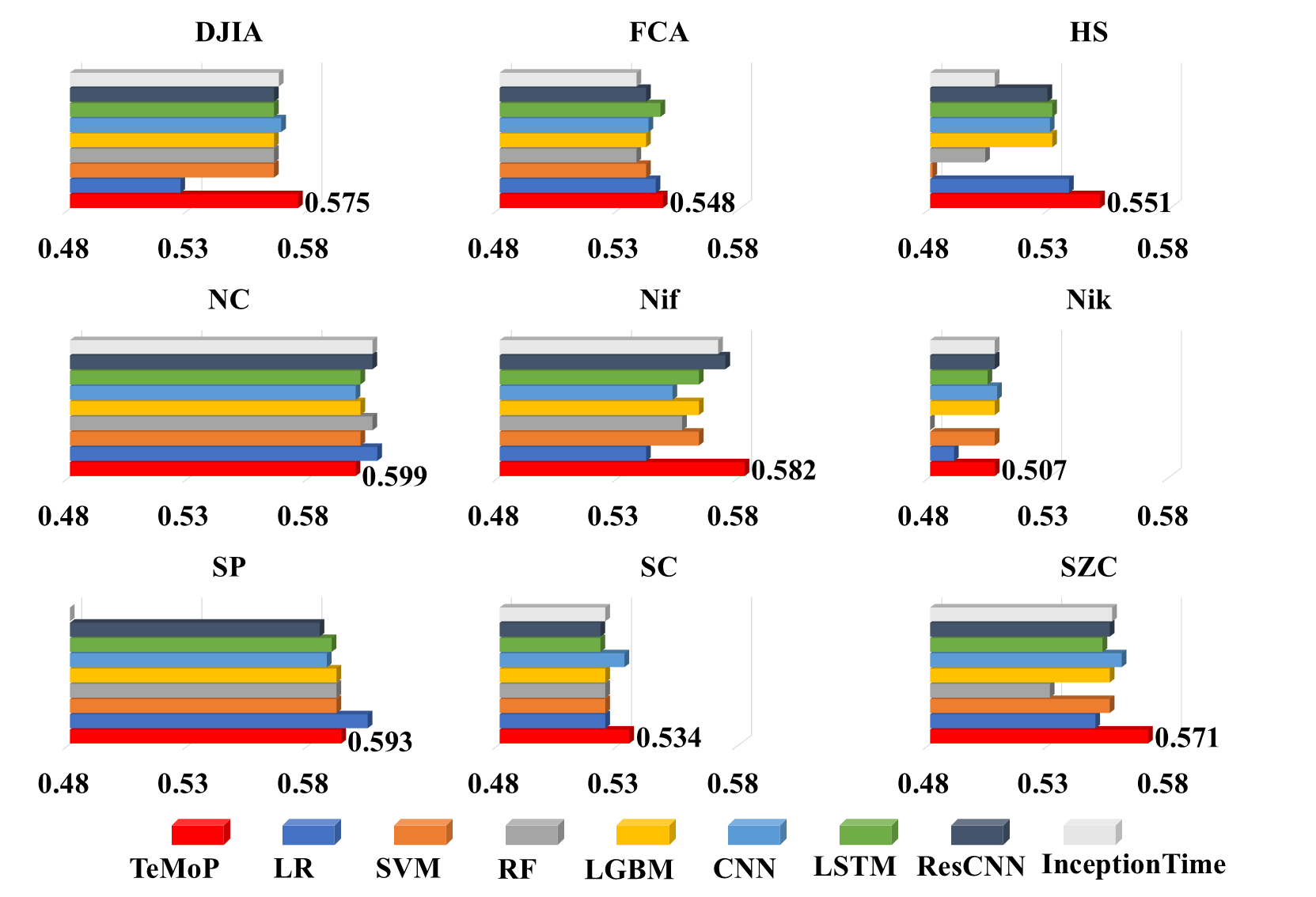}
\caption{ACC of models on different data sets for scheme $2$}\label{fig:scheme2_acc}
\end{figure}

\begin{figure}[htbp!]
\centering
\includegraphics[scale=0.32]{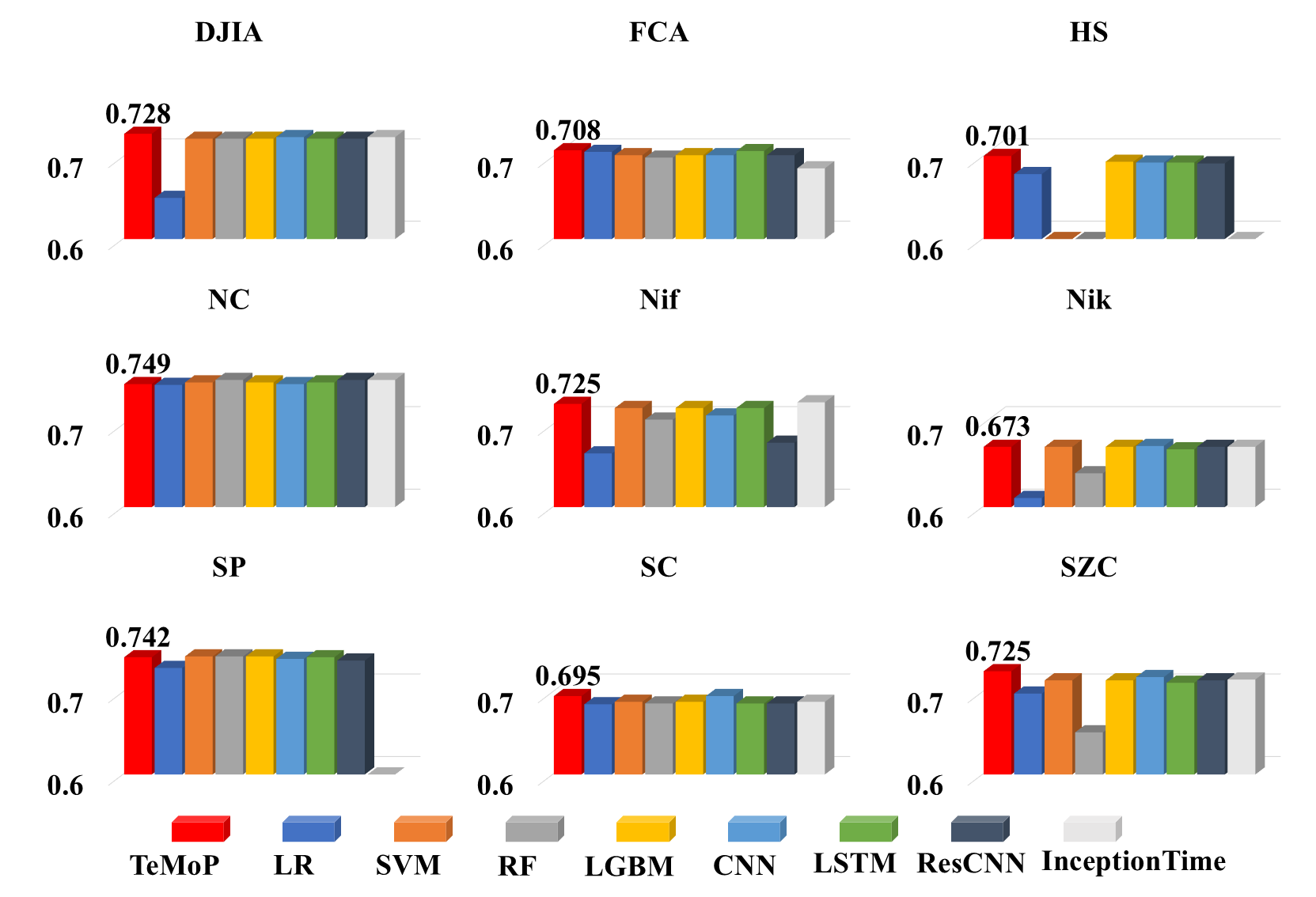}
\caption{F$1$ of models on different data sets for scheme $2$}\label{fig:scheme2_f1}
\end{figure}

\begin{figure}[htbp!]
\centering
\includegraphics[scale=0.32]{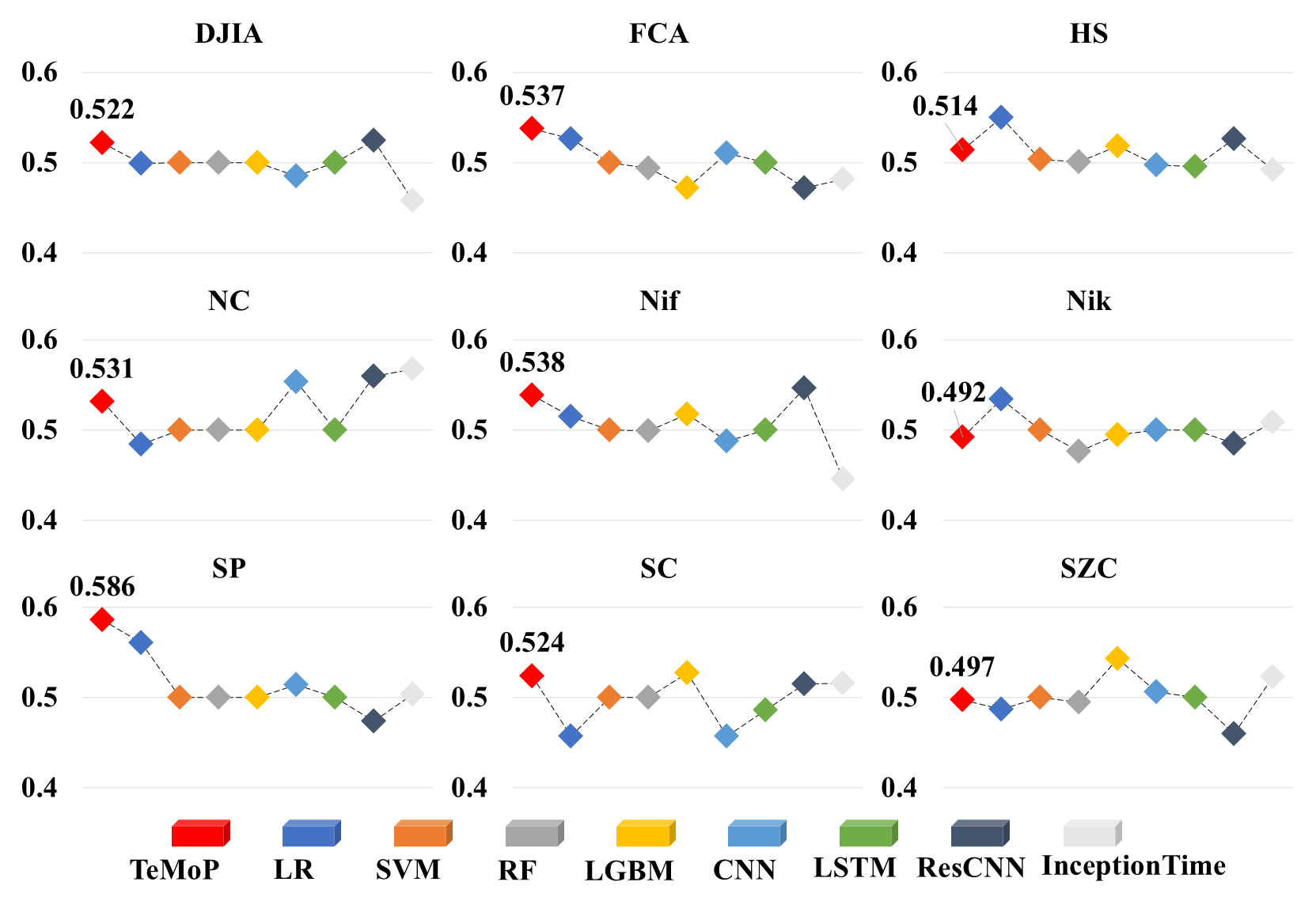}
\caption{AUC of models on different data sets for scheme $2$}\label{fig:scheme2_auc}
\end{figure}

\begin{figure}[htbp!]
\centering
\includegraphics[scale=0.32]{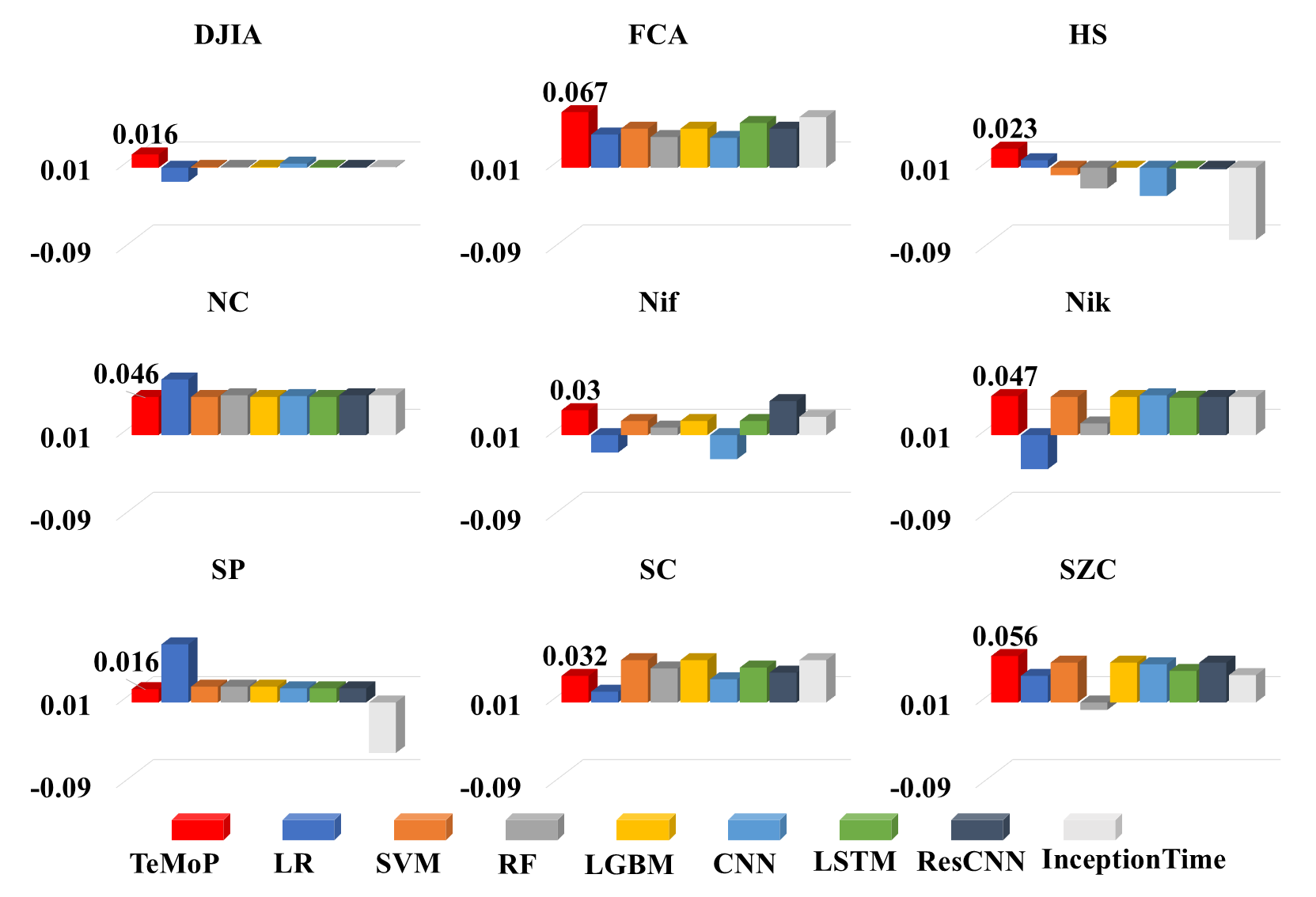}
\caption{SR of models on different data sets for scheme $2$}\label{fig:scheme2_sr}
\end{figure}

Finally, the predictive performance of all models based on this set of hyper-parameters and these lag orders is evaluated on Test using different metrics. Figure \ref{fig:scheme2_acc}, \ref{fig:scheme2_f1}, \ref{fig:scheme2_auc}, \ref{fig:scheme2_sr} show ACC, AUC, F$1$, SR of models on different data sets for scheme $2$ respectively. Specific numerical values are detailed in Table \ref{tab:appendix_acc2}, \ref{tab:appendix_f12}, \ref{tab:appendix_auc2}, \ref{tab:appendix_sr2}. Table \ref{tab:all_metrics2} shows the combined prediction performance of models on all data sets.

\begin{table}[htbp!]
    \centering
    \footnotesize
    \setlength\tabcolsep{1pt}
    \caption{ACC of models on different data sets for scheme $2$}\label{tab:appendix_acc2}
    \begin{tabular}{lccccccccccc}
    \toprule
        model & DJIA & FCA & HS & NC & Nif & Nik & SP & SC & SZC & mean & std \\
    \midrule
        TeMoP & 0.575 &	0.548 &	0.551 &	0.599 &	0.582 &	0.507 &	0.593 &	0.534 &	0.571 & 0.562 & 0.030  \\ 
        LR & 0.526  & 0.545  & 0.538  & 0.608  & 0.541  & 0.490  & 0.604  & 0.524  & 0.549  & 0.547  & 0.038  \\ 
        SVM & 0.565  & 0.541  & 0.481  & 0.601  & 0.563  & 0.507  & 0.591  & 0.524  & 0.555  & 0.548  & 0.039  \\ 
        RF & 0.565  & 0.537  & 0.503  & 0.606  & 0.556  & 0.479  & 0.591  & 0.524  & 0.530  & 0.543  & 0.041  \\ 
        LGBM & 0.565  & 0.541  & 0.531  & 0.601  & 0.563  & 0.507  & 0.591  & 0.524  & 0.555  & 0.553  & 0.031  \\ 
        CNN & 0.568  & 0.542  & 0.530  & 0.599  & 0.552  & 0.508  & 0.587  & 0.532  & 0.560  & 0.553  & 0.029  \\ 
        LSTM & 0.565  & 0.547  & 0.531  & 0.601  & 0.563  & 0.504  & 0.589  & 0.522  & 0.552  & 0.553  & 0.031  \\ 
        ResCNN & 0.565  & 0.541  & 0.529  & 0.606  & 0.574  & 0.507  & 0.584  & 0.522  & 0.555  & 0.554  & 0.032  \\ 
        InceptionTime  & 0.567  & 0.537  & 0.507  & 0.606  & 0.571  & 0.507  & 0.397  & 0.524  & 0.556  & 0.530  & 0.059 \\
    \bottomrule
    \end{tabular}
\end{table}

In Figure \ref{fig:scheme2_acc} and Table \ref{tab:appendix_acc2}, TeMoP has the largest ACC on the remaining data sets except for data set NC, SP. There is a $1.5\%$ difference for the ACC of TeMoP $0.599$ compared to the maximum ACC $0.608$ on data set NC. On data set SP, compared to the maximum ACC $0.604$, the ACC of TeMoP $0.593$ differs by $1.8\%$. The calculations in Table \ref{tab:all_metrics2} show that not only does TeMoP still have the largest mean ACC $0.562$, but TeMoP also has the smallest standard deviation ACC $0.030$. Therefore, for scheme $2$, TeMoP exhibits the smallest prediction error and the best robustness across different data sets when measuring the prediction performance of the model based on the metric ACC.

\begin{table}[htbp!]
    \centering
    \footnotesize
    \setlength\tabcolsep{1pt}
    \caption{F$1$ of models on different data sets for scheme $2$}\label{tab:appendix_f12}
    \begin{tabular}{lccccccccccc}
    \toprule
        model & DJIA & FCA & HS & NC & Nif & Nik & SP & SC & SZC & mean & std \\
    \midrule
        TeMoP & 0.728 &	0.708 &	0.701 &	0.749 &	0.725 &	0.673 	& 0.742 &	0.695 &	0.725 &	0.716 &	0.024  \\ 
        LR & 0.650  & 0.706  & 0.679  & 0.748  & 0.665  & 0.611  & 0.729  & 0.685  & 0.698  & 0.686  & 0.041  \\ 
        SVM & 0.722  & 0.702  & 0.260  & 0.751  & 0.720  & 0.673  & 0.743  & 0.688  & 0.714  & 0.664  & 0.153  \\ 
        RF & 0.722  & 0.699  & 0.536  & 0.754  & 0.706  & 0.641  & 0.743  & 0.686  & 0.651  & 0.682  & 0.066  \\ 
        LGBM & 0.722  & 0.702  & 0.694  & 0.751  & 0.720  & 0.673  & 0.743  & 0.688  & 0.714  & 0.712  & 0.026  \\ 
        CNN & 0.724  & 0.702  & 0.693  & 0.749  & 0.711  & 0.674  & 0.740  & 0.695  & 0.718  & 0.712  & 0.024  \\ 
        LSTM & 0.722  & 0.707  & 0.693  & 0.751  & 0.720  & 0.670  & 0.742  & 0.686  & 0.711  & 0.711  & 0.026  \\ 
        ResCNN & 0.722  & 0.702  & 0.692  & 0.754  & 0.678  & 0.673  & 0.738  & 0.686  & 0.714  & 0.706  & 0.028  \\ 
        InceptionTime  & 0.724  & 0.686  & 0.495  & 0.754  & 0.727  & 0.673  & 0.107  & 0.688  & 0.715  & 0.619  & 0.206 \\
    \bottomrule
    \end{tabular}
\end{table}

Further, Figure \ref{fig:scheme2_f1} and Table \ref{tab:appendix_f12} show the F$1$ scores of the different models. Unlike the case in Scheme $1$ where the F$1$ of the models appeared to be equal to $0$, in Figure \ref{fig:scheme2_f1}, the F$1$ of each model is much larger than $0$, which indicates that the models trained according to Scheme $2$ have better prediction accuracy and improved robustness compared to Scheme $1$. Although the F$1$ scores of all models are very close, TeMoP is still the best performing model among all models. As can be seen from Table \ref{tab:all_metrics2}, it has the largest mean $0.716$ and the smallest standard deviation $0.024$.

\begin{table}[htbp!]
    \centering
    \footnotesize
    \setlength\tabcolsep{1pt}
    \caption{AUC of models on different data sets for scheme $2$}\label{tab:appendix_auc2}
    \begin{tabular}{lccccccccccc}
    \toprule
        model & DJIA & FCA & HS & NC & Nif & Nik & SP & SC & SZC & mean & std \\
    \midrule
        TeMoP & 0.522  & 0.537  & 0.514  & 0.531  & 0.538  & 0.492  & 0.586  & 0.524  & 0.497  & 0.527  & 0.027  \\ 
        LR & 0.499  & 0.526  & 0.550  & 0.484  & 0.515  & 0.534  & 0.560  & 0.457  & 0.487  & 0.512  & 0.034  \\ 
        SVM & 0.500  & 0.500  & 0.503  & 0.500  & 0.500  & 0.500  & 0.500  & 0.500  & 0.500  & 0.500  & 0.001  \\ 
        RF & 0.500  & 0.494  & 0.501  & 0.500  & 0.499  & 0.476  & 0.500  & 0.500  & 0.495  & 0.496  & 0.008  \\ 
        LGBM & 0.500  & 0.472  & 0.518  & 0.500  & 0.517  & 0.495  & 0.500  & 0.527  & 0.543  & 0.508  & 0.021  \\ 
        CNN & 0.485  & 0.510  & 0.497  & 0.553  & 0.488  & 0.500  & 0.514  & 0.457  & 0.506  & 0.501  & 0.026  \\ 
        LSTM & 0.500  & 0.500  & 0.495  & 0.500  & 0.500  & 0.500  & 0.500  & 0.486  & 0.500  & 0.498  & 0.005  \\ 
        ResCNN & 0.524  & 0.472  & 0.526  & 0.559  & 0.546  & 0.485  & 0.474  & 0.515  & 0.460  & 0.507  & 0.035  \\ 
        InceptionTime  & 0.458  & 0.481  & 0.492  & 0.567  & 0.446  & 0.509  & 0.503  & 0.516  & 0.523  & 0.499  & 0.036 \\
    \bottomrule
    \end{tabular}
\end{table}

In terms of the ranking ability of models, except for TeMoP and LR, AUC of the rest of the models is around $0.5$ in Figure \ref{fig:scheme2_auc} and Table \ref{tab:appendix_auc2}, which reflect the poor ranking ability of these models. Except for SZC and Nik, AUC of TeMoP on the rest of the data sets takes values significantly greater than $0.5$. In Table \ref{tab:all_metrics2}, Combining all the data sets, TeMoP has the largest mean $0.527$ and smaller standard deviation $0.027$, which demonstrates the superiority and robustness of TeMoP in terms of model ranking ability.

\begin{table}[htbp!]
    \centering
    \footnotesize
    \setlength\tabcolsep{1pt}
    \caption{SR of models on different data sets for scheme $2$}\label{tab:appendix_sr2}
    \begin{tabular}{lccccccccccc}
    \toprule
        model & DJIA & FCA & HS & NC & Nif & Nik & SP & SC & SZC & mean & std \\ 
    \midrule
        TeMoP & 0.016 &	0.067 &	0.023 &	0.046 &	0.030 &	0.047 &	0.016 &	0.032 &	0.056 &	0.037 &	0.018  \\ 
        LR & -0.017  & 0.040  & 0.009  & 0.067  & -0.021  & -0.041  & 0.070  & 0.013  & 0.032  & 0.017  & 0.039  \\ 
        SVM & 0.001  & 0.047  & -0.009  & 0.046  & 0.017  & 0.046  & 0.019  & 0.051  & 0.048  & 0.030  & 0.023  \\ 
        RF & 0.001  & 0.037  & -0.025  & 0.048  & 0.009  & 0.014  & 0.019  & 0.041  & -0.009  & 0.015  & 0.024  \\
        LGBM & 0.001  & 0.047  & 0.000  & 0.046  & 0.017  & 0.046  & 0.019  & 0.051  & 0.048  & 0.031  & 0.021  \\ 
        CNN & 0.005  & 0.036  & -0.034  & 0.047  & -0.029  & 0.048  & 0.017  & 0.028  & 0.046  & 0.018  & 0.032  \\ 
        LSTM & 0.001  & 0.054  & -0.001  & 0.046  & 0.017  & 0.045  & 0.017  & 0.042  & 0.038  & 0.029  & 0.021  \\ 
        ResCNN & 0.001  & 0.047  & -0.002  & 0.048  & 0.041  & 0.046  & 0.017  & 0.036  & 0.048  & 0.031  & 0.021  \\ 
        InceptionTime  & 0.001  & 0.061  & -0.087  & 0.048  & 0.022  & 0.046  & -0.061  & 0.051  & 0.033  & 0.013  & 0.053 \\ 
    \bottomrule
    \end{tabular}
\end{table}

In terms of simulated returns, although all models take positive values on mean as shown in Table \ref{tab:all_metrics2}, only the LGBM and TeMoP take SR values greater than $0$ on all data sets as shown in Figure \ref{fig:scheme2_sr} and Table \ref{tab:appendix_sr2}, reflecting the superiority and robustness of the LGBM and TeMoP in terms of simulated returns. Among all models, TeMoP has the largest mean SR $0.037$ and the smallest standard deviation $0.018$, which TeMoP demonstrates superiority and robustness in simulating returns compared to other models.

\begin{table}[htbp!]
    \centering
    \footnotesize
     \caption{Prediction performance of models for scheme $2$}\label{tab:all_metrics2}
    \begin{tabular}{lcccc}
    \toprule
        model & ACC & F$1$ & AUC & SR \\ 
    \midrule
        TeMoP & \textbf{0.562$\pm$0.030}$^{a}$ & \textbf{0.716$\pm$0.024} & \textbf{0.527$\pm$0.027} & \textbf{0.037$\pm$0.018}\\
        LR & 0.547$\pm$0.038 & 0.686$\pm$0.041 & 0.512$\pm$0.034 & 0.017$\pm$0.039\\
        SVM & 0.548$\pm$0.039 & 0.664$\pm$0.153 & 0.500$\pm$0.001 & 0.030$\pm$0.023\\
        RF & 0.543$\pm$0.041 & 0.682$\pm$0.066 & 0.496$\pm$0.008 & 0.015$\pm$0.024\\
        LGBM & 0.553$\pm$0.031 & 0.712$\pm$0.026 & 0.508$\pm$0.021 & 0.031$\pm$0.021\\
        CNN & 0.553$\pm$0.029 & 0.712$\pm$0.024 & 0.501$\pm$0.026 & 0.018$\pm$0.032\\
        LSTM & 0.553$\pm$0.031 & 0.711$\pm$0.026 & 0.498$\pm$0.005 & 0.029$\pm$0.021 \\
        ResCNN & 0.554$\pm$0.032 & 0.706$\pm$0.028 & 0.507$\pm$0.035 & 0.031$\pm$0.021\\
        InceptionTime & 0.530$\pm$0.059 & 0.619$\pm$0.206 & 0.499$\pm$0.036 & 0.013$\pm$0.053\\
    \bottomrule
    \end{tabular}
    \begin{flushleft}
         $^{a}$ 0.562$\pm$0.030 denotes mean of this metric $\pm$ standard deviation of this metric between different data sets. $^{b}$ Bold indicates that this result is optimal for all models.
    \end{flushleft}
\end{table}

Combining the results of all the data sets we find that similar to Scheme $1$, TeMoP shows significant superiority and robustness in terms of prediction accuracy, model ranking ability, and simulated returns. In addition, we also find that the way in which the lag order is chosen as a hyper-parameter based on the models' performance on the data sets has more robust results compared to the way in which it is specified directly.

\section{Summary and Outlook}\label{sec:sum}

Based on the unsatisfactory performance of existing models in dealing with time series trend problems, this paper designs a novel multiple lag order probabilistic model based on trend encoding TeMoP. Traditional forecasting models often use samples of a certain lag order for training and forecasting. However, this approach tends to make the prediction model poorly robust on different data sets, or on different time periods of the same data set. Therefore, from the design principle, TeMoP discards this approach. 

TeMoP adaptively calculates the maximum lag order based on the nature of data. Then, for each lag order $i$ that does not exceed the maximum lag order, it obtains a sample set $\Omega_i$, and trains a model based on $\Omega_i$. Finally, the calculation results of each model are considered comprehensively in order to improve robustness on different data sets. In addition, TeMoP introduces trend features to improve the predictive accuracy.

In order to comprehensively analyze the forecasting performance of TeMoP on realistic data sets, a total of nine stock index data sets from three different market types are selected for the experiment. There are three data sets from developed markets, two data sets from semi-developed markets and four data sets from developing markets. Meanwhile, the experiment selects a total of six classical models from three different domains, namely statistics, narrow machine learning, and deep learning, for comparative analysis with TeMoP. In addition, two state-of-the-art models are added to the experiment to compare the advantages and disadvantages between TeMoP and state-of-the-art models.

Unlike the comparison models, TeMoP not only can adaptively select the maximum lag order, but also is a non-parametric model. Therefore, TeMoP avoids the step of choosing the optimal hyper-parameters. Therefore, in order to enhance the fairness of the comparison between different models, two comparison schemes are designed in the experiment. Scheme $1$ specifies the lag order, and Scheme $2$ treats the lag order as a hyper-parameter of models chosen based on models' performance on data.

Combining the experimental results of Scheme $1$ and $2$, we find that:
\begin{itemize}
    \item TeMoP shows significant superiority compared to the comparison models, both in terms of prediction accuracy, model ordering ability and simulation gains based on model prediction results. At the same time, this superiority shows good robustness across different data sets.
    \item The predictive performance of the comparison model fluctuates greatly between different data sets, and also fluctuates greatly between different lag orders. This indicates that the robustness of the comparison model is poor.
\end{itemize}

This study mainly analyzes the prediction of the next trend of TeMoP. However, in practical applications, the prediction of $n$ steps under time series data is also of great significance. Therefore, the future work will focus on the model performance of TeMoP in the next $n$ step prediction to enhance the practical application significance.


 \bibliography{proba-cas-refs}
 \bibliographystyle{chicago}
 \biboptions{authoryear}

 \end{document}